\documentclass[3p,review]{elsarticle}
\usepackage{geometry}                
\usepackage{graphicx}
\usepackage{subfig}
\usepackage{amssymb}
\usepackage{amsmath}
\usepackage{epstopdf}
\usepackage{color}
\usepackage{bm}
\usepackage{mathrsfs} 
\usepackage{hyperref}
\usepackage{wrapfig}
\usepackage{textcomp}
\usepackage{tabularx}
\usepackage{comment}
\usepackage{subeqnarray}
\usepackage{mathtools}
\usepackage{multirow}
\usepackage{placeins}
\newcolumntype{Y}{>{\centering\arraybackslash}X}

\usepackage[nameinlink,noabbrev]{cleveref}
\makeatletter
\def\@seccntformat#1{\@ifundefined{#1@cntformat}%
   {\csname the#1\endcsname\quad}
   {\csname #1@cntformat\endcsname}
}
\makeatother
\usepackage{scalerel}[2014/03/10]
\usepackage{stackengine}

\begin{document}

\begin{frontmatter}
 \title{ An explicit algebraic closure for passive scalar-flux: Applications in heated 
channel flows subjected to system rotation 
\footnote{This research did not receive any specific grant from funding agencies in the 
public, commercial, or not-for-profit sectors.} }
\author[1]{C.~F.~Panagiotou\corref{cor1}}
\author[2]{F.~S.~Stylianou}
\author[3,4]{E.~Gravanis}
\author[3,4]{E.~Akylas}
\author[2]{S.~C.~Kassinos}
\address[1]{ Nireas International Water Research Center,\\
 Department of Civil \& Environmental Engineering,\\
 University of Cyprus, 75 Kallipoleos, Nicosia 1678, Cyprus }
\address[2]{ Computational Sciences Laboratory, UCY-CompSci,\\
 Department of Mechanical \& Manufacturing Engineering,\\
 University of Cyprus, 75 Kallipoleos, Nicosia 1678, Cyprus }
\address[3]{ Department of Civil Engineering \& Geomatics,\\
Cyprus University of Technology, PO Box 50329, Limassol 3603, Cyprus }
\address[4]{ Eratosthenes Centre of Excellence,\\
Cyprus University of Technology, Limassol, Cyprus }
\cortext[cor1]{Corresponding author: Constantinos Panagiotou, Email address: panagiotou.konstantinos@ucy.ac.cy}

\begin{abstract}
We present an algebraic model for turbulent scalar-flux vector that 
stems from tensor representation theory. The resulting closure  
contains direct dependence on mean velocity gradients and on frame rotation tensor that
accounts for Coriolis effects.
Model coefficients are determined from Direct Numerical Simulations (DNS) data of homogeneous
shear flows subjected to arbitrary mean scalar gradient orientations. This type of tuning
process renders the proposed model to be objective towards inhomogeneous applications.
Model performance is evaluated in several heated channel flows in both stationary and 
rotating frames, showing good results. To place the performance of the proposed model into
context, we compare with Younis algebraic model \cite{Younis2010}, which is known to provide 
reasonable predictions for several engineering flows.
\end{abstract}

\end{frontmatter}

\pagestyle{plain}

\newcount\ndots
\def\drawline#1#2{\raise 2.5pt\vbox{\hrule width #1pt height #2pt}}
\def\spacce#1{\hskip #1pt}
\def\solid{\drawline{24}{.5}\nobreak\ }
\def\bdash{\hbox{\drawline{4}{.5}\spacce{2}}}
\def\dashed{\bdash\bdash\bdash\bdash\nobreak\ }
\def\bdot{\hbox{\drawline{1}{.5}\spacce{2}}}
\def\dotted{\hbox{\leaders\bdot\hskip 24pt}\nobreak\ }
\def\chndash{\hbox {\drawline{8.5}{.5}\spacce{2}\drawline{3}{.5}\spacce{2}\drawline{8.5}{.5}}\nobreak\ }
\def\chndot{\hbox {\drawline{9.5}{.5}\spacce{2}\drawline{1}{.5}\spacce{2}\drawline{9.5}{.5}}\nobreak\ }
\def\chndotdot{\hbox {\drawline{8}{.5}\spacce{2}\drawline{1}{.5}\spacce{2}\drawline{1}{.5}\spacce{2}\drawline{8}{.5}}\nobreak\ }
\def\chndotdotdot{\hbox {\drawline{8}{.5}\spacce{2}\drawline{1}{.5}\spacce{2}\drawline{1}{.5}\spacce{2}\drawline{1}{.5}\spacce{2}\drawline{8}{.5}}\nobreak\ }
\def\trian{\raise 1.25pt\hbox{$\scriptscriptstyle\triangle$}\nobreak\ }
\def\circle{$\circ$\nobreak\ }
\def\diam{$\diamond$\nobreak\ }
\def\solidcircle{$\bullet$\nobreak\ }

\def\smalltriangle{$\scriptstyle\triangle\textstyle$\nobreak\ }
\def\smallplus{$\scriptstyle + \textstyle$\nobreak\ }
\def\smalltimes{$\scriptstyle\times\textstyle$\nobreak\ }
\def\smallnabla{$\scriptstyle\nabla\textstyle$\nobreak\ }
\def\square{${\vcenter{\hrule height .4pt
        \hbox{\vrule width .4pt height 3pt \kern 3pt
        \vrule width .4pt}
        \hrule height .4pt}}$\nobreak\ }
\def\plus{\raise 1.25pt \hbox{$\scriptscriptstyle +$}\nobreak\ }
\def\x{\raise 1.25pt \hbox{$\scriptscriptstyle \times$}\nobreak\ }
\def\ldash{\hbox {\drawline{7}{.5}\spacce{2}\drawline{7}{.5}\spacce{2}\drawline{7}{.5}}\nobreak\ }
\def\lchndash{\hbox {\drawline{15}{.5}\spacce{3}\drawline{7}{.5}}\nobreak\ }
\def\tsolid{\drawline{24}{1.2}\nobreak\ }

\def\graytrian{\raise 1.25pt
   \hbox to 3bp{
\hfill}\nobreak\ }
\def\trian{\raise 1.25pt
   \hbox to 3bp{
\hfill}\nobreak\ }
\def\solidtrian{\raise 1.25pt
   \hbox to 3bp{
\hfill}\nobreak\ }

\def\graytriand{\raise 1.25pt
   \hbox to 3bp{
\hfill}\nobreak\ }
\def\triand{\raise 1.25pt
   \hbox to 3bp{
\hfill}\nobreak\ }
\def\solidtriand{\raise 1.25pt
   \hbox to 3bp{
\hfill}\nobreak\ }

\def\square{\raise 1.pt
   \hbox to 3bp{
\hfill}\nobreak\ }
\def\graysquare{\raise 1.pt
   \hbox to 3bp{
\hfill}\nobreak\ }
\def\solidsquare{\raise 1.pt
   \hbox to 3bp{
\hfill}\nobreak\ }

\def\circle{\raise 1.pt
   \hbox to 3bp{
\hfill}\nobreak\ }
\def\solidcircle{\raise 1.pt
   \hbox to 3bp{
\hfill}\nobreak\ }
\def\graycircle{\raise 1.pt
   \hbox to 3bp{
\hfill}\nobreak\ }

\def\dotcirc{$\cdots\ $\circle$\cdots$\ }

\def\dashx {\bdash\bdash\smalltimes\bdash\bdash}

\def\chndashx {\drawline{8.5}{.5}\spacce{2}
\drawline{3}{.5}$\scriptstyle\times\textstyle$\drawline{8.5}{.5}\spacce{2}
\drawline{3}{.5}\nobreak\ }

\def\solidcclose{\drawline{10}{.5}\nobreak\raise
  0.5pt\hbox{$\bullet$}\drawline{10}{.5}\nobreak\ }

\def\solidsclose{\drawline{10}{.5}\nobreak\raise
  0.5pt\hbox{\solidsquare}\drawline{10}{.5}\nobreak\ }

\def\solidtclose{\drawline{10}{.5}\nobreak\raise
  0.5pt\hbox{\solidtrian}\drawline{10}{.5}\nobreak\ }

\def\solidcopen{\drawline{10}{.5}\nobreak\raise
  0.5pt\hbox{\circle}\drawline{10}{.5}\nobreak\ }

\def\solidsopen{\drawline{10}{.5}\nobreak\raise
  0.5pt\hbox{\square}\drawline{10}{.5}\nobreak\ }

\def\solidtopen{\drawline{10}{.5}\nobreak\raise
  0.5pt\hbox{\trian}\drawline{10}{.5}\nobreak\ }

\def\solidx{\drawline{10}{.5}\nobreak\raise
  0.5pt\hbox{\x}\drawline{10}{.5}\nobreak\ }

  
\font\msakkk=msam10
\def\diamsol{{\msakkk \char7}}
\def\diamop{{\msakkk \char6}}
\def\starsol{{\msakkk \char70}}
\def\triansolu{{\msakkk \char78}}
\def\triansold{{\msakkk \char72}}
\def\triansolr{{\msakkk \char73}}
\def\triansoll{{\msakkk \char74}}
\def\trianopu{{\msakkk \char77}}
\def\trianopd{{\msakkk \char79}}
\def\trianopr{{\msakkk \char66}}
\def\trianopl{{\msakkk \char67}}
\def\squarsol{{\msakkk \char4}}
\def\squarop{{\msakkk \char3}}

\def\mydash{\hbox{\drawline{2}{.5}\spacce{2}}}
\def\shdashed{\mydash\mydash\mydash\mydash\mydash\mydash\nobreak\ }

\def\bdot{\hbox{\drawline{.5}{.5}\spacce{1}}}
\def\dotted{\hbox{\leaders\bdot\hskip 24pt}\nobreak\ }

\newcommand{\AC}[1]{\vspace{.5cm} [A COMPLETER: #1]}
\newcommand{\CTRClass}{{\it{\bf ctr\_summer.cls }}}
\newcommand{\noi}{\par}

\section{INTRODUCTION}\label{introduction}

Turbulent scalar-flux vector has a notable role in the scalar transport
for a wide range of practical applications. It appears in the Reynolds-Averaged 
scalar transport equations as a term that needs to be modelled so that closure is achieved.
An elegant choice is the development of engineering models with aim to provide 
estimations of this quantity at low computational complexity.  Engineering models for the 
scalar-flux are classified into two categories: differential and algebraic.
Differential transport models (DTM) are proven to be beneficial tools, being capable of 
handling rotational and curvature effects. However, robustness issues and performance inconsistencies, combined  
with computational overheads associated with the solution of a 
differential transport equation for each scalar-flux component, 
prevent this class of models from penetrating further into the mainstream of engineering 
practice.
On the other hand, algebraic approaches are based on assumptions that lead to constitutive
equations between turbulent statistics and  mean deformation. The simplest algebraic model is based on the 
gradient-diffusion hypothesis (GDH), which assumes that turbulent scalar-flux is aligned to the 
mean scalar gradient. 
Despite its implementational and computational elegance, this assumption is an
important reason why this model fails to capture important flow features, such as 
turbulence anisotropy or the effects of mean and system rotation. As a result, 
Batchelor \cite{Batchelor1949} proposed a generalization of GDH 
(GGDH)  from which several algebraic closures have emerged.
For example, Daly $\&$ Harlow \cite{Daly1970} adopted  scale functions to express 
the turbulent scalar-flux vector as a product between Reynolds stress and mean scalar-gradient, 
allowing  the mis-alignment of the scalar-fluxes with the mean scalar deformation. However, 
this expression is found to perform poorly on predicting the correct magnitudes of 
scalar-flux components in the directions normal to the mean scalar gradient.  
Suga $\&$ Abe \cite{Suga2000} improved the performance of the GGDH model by adding a non-linear 
term  that contains quadratic products of the Reynolds stress tensor, which correctly captures 
the anisotropy levels of the scalar-flux components in the vicinity of wall boundaries. 
This addition resulted in the construction of Higher-Order GGDH models (HO-GGDH), which were successfully applied 
in a wide range of heated channel flows. 
Another interesting approach is to derive algebraic expressions directly from the 
exact transport equation of turbulent scalar-flux through equilibrium assumptions. 
A common choice is to apply the weak-equilibrium assumption (WEA) \cite{Rodi1972}, which 
states that the transient variations of turbulent anisotropies are negligible compared to 
the variation of turbulent scales. To some extent, these models are considered to be a good alternative 
to DTM, as they have been successfully applied in different flow configurations while 
requiring less computational capacity \cite{Girimaji1998, Wikstrom2000, Lazeroms2013}.   

An alternative approach for estimating the turbulent scalar-fluxes was proposed by
Younis et al. \cite{Younis2005}.
Using as guidance the exact scalar-flux transport equation, Younis and coworkers 
expressed this quantity as a function of several tensor quantities. 
This approach is elegant, since it avoids the reduction of  
transport equations through the WEA with all the modelling uncertainties that entails. It also
provides a general framework, from which different algebraic expressions can be obtained.
They proposed a multi-linear closure that  
exhibited distinct improvements over other algebraic scalar-flux closures in benchmark 
two-dimensional  free shear flows, while it was successfully used in flow
configurations involving Coriolis and curvature effects \cite{Younis2007,Muller2015}. 
However, the linear nature of this specific closure is the main reason why it fails to capture the 
proper near-wall anisotropy levels of scalar-flux vector, thus revealing the importance of incorporating 
non-linear information regarding turbulence anisotropy.
Hence, in this study we propose an algebraic model that stems directly from the Younis formulation 
and involves products of the Reynolds stress tensor, a dependence missing from the 
multi-linear model. Special attention is given so that model complexity is kept minimal for 
ease implementation in existing industrial codes, while its performance ability is tested 
on several heated Couette and Poiseuille flows in stationary and rotating frames.

A description of the exact transport equations of the turbulent passive-scalar fluxes along with
an extensive summary of the Younis approach are given in Section~\ref{younis_approach} with 
the motivation behind the use of a non-linear term involving products of Reynolds stress  
given in Section~\ref{NLTERM_INCLUSION}. The proposed formulation is discussed in 
Section~\ref{proposed_formulation} and the performance of this closure is evaluated in 
Section~\ref{MODEL_ASSES}, yielding good results. Summary and conclusions are 
given in Section~\ref{SUMMARY_AND_CONCLUSIONS}.

\section{ YOUNIS' FORMULATION }\label{younis_approach}

The motivation behind the work reported  by Younis et al. \cite{Younis2005} arose 
from the need to provide a better alternative to the existing gradient-transport closures 
for the turbulent scalar-fluxes. 
As starting point, they considered the exact transport equation of these fluxes in 
a frame rotating at a constant angular velocity rate $\Omega^{f}$
\begin{equation}\label{scalar_flux_exact_equation}
\begin{split}
\frac{\partial \overline{u'_{i}\phi'} }{\partial t} + 
\overline{u}_{j}\frac{\partial \overline{u'_{i}\phi'} }{\partial x_{j}} = &
-\overline{u'_{i}u'_{j}}\frac{ \partial \overline{\phi} }{ \partial x_{j}}
-\overline{u'_{j}\phi'}\frac{\partial \overline{u}_{i}}{\partial x_{j}} 
- 2 \epsilon_{ijk} \Omega^{f}_{j} \overline{u'_{k}\phi'} \\
& - \frac{1}{\rho} \overline{ p'\frac{\partial \phi' }{ \partial x_{i} } } 
- (\gamma + \nu) \overline{ \frac{\partial \phi' }{ \partial x_{j} } \frac{\partial u'_{i} }{ \partial x_{j} }  } \\
& -\frac{\partial}{\partial x_{j}}\bigg( \overline{u'_{i}u'_{j}\phi'} 
+ \frac{1}{\rho} \overline{p'\phi'}\delta_{ij}-\gamma \overline{u'_{i}\frac{\partial \phi'}{\partial x_{j}} }
-\nu \overline{ \phi'\frac{\partial u'_{i}}{\partial x_{j}} }\bigg)\,,
\end{split}
\end{equation}
where the instantaneous flow variables are decomposed into a mean part, denoted by overbar, 
and a fluctuating part, denoted by prime symbol.  
Hereafter, we are using index notation whereby repeated indexes imply summation. The 
coefficient of fluid viscosity and the 
coefficient of scalar  diffusivity are denoted by $\nu$ and  $\gamma$ respectively, 
$\rho$ is the density of the fluid  and $u_i$, $\phi$ are the instantaneous fluid velocity 
and passive scalar fields respectively. The fluctuating pressure is denoted with $p'$.
The first and second terms on the RHS of equation~\eqref{scalar_flux_exact_equation} represent 
the generation of scalar-flux as a consequence of turbulence-mean deformation interactions.
The third term arises when flow is subjected to system rotation around an arbitrary axis,
while the fourth term refers to the fluctuating pressure-scalar correlations and is responsible
for the  redistribution of the flux among the different components.
The fifth term refers to the rate at which the scalar-flux is destructed, while the last
term is interpreted as the turbulent-transport term.
Younis et al. \cite{Younis2005} used equation~\eqref{scalar_flux_exact_equation} as a guide 
to provide a rationally assumed relationship between the scalar-flux vector and various 
tensor quantities,
\begin{equation}\label{functional_relationship}
\overline{ u'_{i}\phi'  } = f_{i}\bigg(  R_{ij},S_{ij},W_{ij},\Omega^{f}_{ij}, 
\Lambda_{i},\epsilon, \epsilon_{\phi},\overline{ \phi^{'2} }  \bigg) \,, 
\end{equation}
where $R_{ij}=\overline{u'_i u'_j }$ is the Reynolds stress tensor, 
$ \epsilon $ is the energy dissipation rate,
$\epsilon_{\phi}$ is half the scalar dissipation rate and $\overline{ \phi^{'2} }$ is the scalar-variance. 
$S_{ij}$, $W_{ij}$, $\Omega^{f}_{ij}$ and $\Lambda_{i}$ denote the mean strain-rate tensor,
the mean vorticity tensor, the frame-rotation rate tensor and the mean scalar gradient vector 
respectively, defined as
\begin{equation}
\begin{split}
S_{ij} & = \frac{1}{2}\bigg( \frac{ \partial \overline{u}_{i} }{ \partial x_{j} } + 
\frac{ \partial \overline{u}_{j} }{ \partial x_{i} }     \bigg)\,,\qquad 
W_{ij} = \frac{1}{2} \bigg( \frac{ \partial \overline{u}_{i} }{ \partial x_{j} } - 
\frac{ \partial \overline{u}_{j} }{ \partial x_{i} } \bigg)\,,\qquad 
\Omega^{f}_{ij} = \epsilon_{ikj}\, \Omega^{f}_{k}\,, \qquad
\Lambda_{i} = \frac{ \partial \overline{\phi} }{ \partial x_{i} }\,. 
\end{split}
\end{equation} 
With the aid of tensor representation theory, Younis et al. constructed an explicit model for 
$\overline{ u'_{i}\phi' }$ in accordance with equation~\eqref{functional_relationship}. 
Under this approach,  $\overline{u'_{i}\phi'}$ can be expressed as a 
series of basis vectors $\Upsilon_{i}$,
\begin{equation}
\overline{ u'_{i}\phi' } = \sum^{M}_{n=1} \alpha_{n} \Upsilon^{n}_{i}\,,
\end{equation}
where $\alpha_{n}$ can depend on all the tensor variables appearing in 
equation~\eqref{functional_relationship}. 
The basis vectors  are formed from the products of the symmetric ($S_{ij}, R_{ij}$), 
the skew-symmetric ($W_{ij}$) tensors  and the vector ($\Lambda_{i}$), 
leading to the following algebraic expression
\begin{equation}\label{general_representation}
\begin{split}
\overline{ u'_{i}\phi' }&= \alpha_{1}\Lambda_{i} 
					     + \alpha_{2} R_{ij}\Lambda_{j} 
					     + \alpha_{3} S_{ij}\Lambda_{j}
					     + \alpha_{4} R_{ik}R_{kj}\Lambda_{j}
					     + \alpha_{5} S_{ik}S_{kj}\Lambda_{j}
					     + \alpha_{6} W_{ij}\Lambda_{j} \\
					    & 
					     + \alpha_{7} W_{ik}W_{kj}\Lambda_{j}
					     + \alpha_{8} \bigg( S_{ik}W_{kj} + S_{jk}W_{ki} \bigg)\Lambda_{j}
					     + \alpha_{9} \bigg( R_{ik}S_{kj} + R_{jk}S_{ki} \bigg)\Lambda_{j} \\
					    & 
					     + \alpha_{10}\bigg( R_{ik}W_{kj} + R_{jk}W_{ki} \bigg)\Lambda_{j}\,,
\end{split}
\end{equation}
where for simplicity we neglected the presence of Coriolis effects in the above expression.
To bring the above expression into a more compact form, Younis et al. \cite{Younis2005}  assumed  sufficiently small
anisotropies and turbulent time scales. Further assuming that the effects of $S_{ij}$ and 
$W_{ij}$ are in balance, results to the following multilinear expression:
\begin{equation}\label{Younis_expression}
\overline{u'_{i}\phi'} =  C_{1} \frac{\kappa^{2}}{\epsilon}\Lambda_{i}
  				        + C_{2} \frac{\kappa}{\epsilon} R_{ij}\Lambda_{j}
  				        + C_{3} \frac{\kappa^{3}}{\epsilon^{2}}G^{T}_{ij}\Lambda_{j}
  				        + C_{4} \frac{\kappa^{2}}{\epsilon^{2}} \bigg( R_{ik}G^{T}_{jk} + 
  				                 R_{jk}G^{T}_{ik} \bigg)\Lambda_{j}\,,\qquad 
G^{T}_{ij} = G_{ij} + \Omega^{f}_{ij}\,,  				                 
\end{equation}
where $\kappa = R_{kk}/2$ is the turbulent kinetic energy, while the $\alpha_i$ coefficients were 
replaced based on dimensional arguments. Coriolis effects are explicitly incorporated in the 
closure by adding the frame-rotation tensor $\Omega^{f}_{ij}$ into the mean velocity gradient 
tensor $G_{ij}=S_{ij}+W_{ij}$, an approach also followed in previous studies \cite{Launder1987}.
The first term on the RHS of equation~\eqref{Younis_expression} represents the GDH, while 
the second term coincides with the production term that appears in equation~\eqref{scalar_flux_exact_equation} 
and is associated with the mean scalar deformation, 
also known as the Generalized Gradient Diffusion Hypothesis (GGDH) model proposed by Daly 
$\&$ Harlow \cite{Daly1970}.
The remaining two terms involve products between the gradients of mean scalar and mean velocity;
a dependence proposed by the analyses of Dakos $\&$ Gibson~\cite{Dakos1987} and 
Yoshizawa~\cite{Yoshizawa1985}.
The values of $C_{i}$ coefficients were determined based on the LES results of Kaltenbach 
\cite{kaltenbach1994} for homogeneous flows subjected to a uniform shear with uniform 
scalar gradients
\begin{equation}
C_{1} = 0.0455\,,\qquad C_{2} = -0.373 \,,\qquad C_{3} = 0.00373\,,\qquad C_{4} = 0.0235\,.
\end{equation}
Regarding inhomogeneous flows, detailed analysis of the model performance in relation to 
data from a large number of studies on wall-bounded  flows showed a serious error in the 
normal flux component \cite{Younis2007}. As a result, the value of $C_{1}$  coefficient was modified in the 
near-wall region by applying the following damping function
\begin{equation}\label{C1_damping_func}
C_{1} = 0.0455\,f_{C_{1}},\qquad f_{C1}= 1-\exp(-A\,\beta \, Pe^{\alpha}),\qquad 
Pe = Pr\,Re_{\tau}\,,
\end{equation} 
where $ \alpha= -0.02 $ and $\beta=1.9$ are the values proposed by Younis et al. \cite{Younis2010}, 
$Pr=\nu/\gamma$ is the Prandtl number, $Re_{t}=\frac{\kappa^{2}}{\nu \epsilon}$ is the turbulence Reynolds number, while 
$A$ is the stress-flatness parameter, defined as
\begin{equation}
\begin{split}
A_{ij} & = \frac{R_{ij}}{\kappa}-\frac{2}{3}\delta_{ij}\,,\qquad A_{2} = A_{ij}A_{ji}\,,
\qquad A_{3} = A_{ij}A_{jk}A_{ki}\,,\qquad A = 1-\frac{9}{8}(A_{2}-A_{3})\,,
\end{split}
\end{equation}
where $A_{2}$ and $A_{3}$ are the second and third invariants of the normalized Reynolds-stress
anisotropic tensor $A_{ij}$.
More details regarding the mathematical formulation can be found in Younis et al. \cite{Younis2007}.
  
As already mentioned in the introduction, the linear form of Younis closure is an important 
reason why this closure exhibits certain limitations. For example, it fails to capture the 
near-wall anisotropy levels of the scalar-flux vector.
In the following section we discuss the importance of incorporating a term that 
contains non-linear information regarding the turbulent anisotropy, vital to predict the 
proper near-wall behaviour.

\section{QUADRATIC DEPENDENCE ON REYNOLDS STRESS}\label{NLTERM_INCLUSION}

It is well known that models involving only the second term of equation~\eqref{Younis_expression} 
cannot predict the streamwise heat-flux component reasonably well \cite{Launder1988}.
In an attempt to extend their applicability, Abe $\&$ Suga \cite{Abe2001}
performed a series of LES simulations for fully-developed turbulent channel flows under 
different boundary conditions and for a wide range of Prandtl numbers. 
They relied on the findings of Kim $\&$ Moin \cite{Kim1989}, who pointed out that scalar fluctuations are
correlated more strongly with streamwise than transverse velocity fluctuations in the near-wall
region,
to propose an algebraic relation between  the turbulent scalar-flux vector and quadratic 
products of Reynolds stress tensor
\begin{equation}\label{Abe_Suga_Proposal}
\overline{ u'_{i}\phi' } = -C_{\phi} \tau \bigg( \frac{ R_{ik}R_{kj} }{ \kappa }  \bigg) \Lambda_{j}\,,
\end{equation}  
where $\tau$ is a turbulent time scale and $C_{\phi}$ is a model coefficient.
In wall-bounded channel flows where statistical quantities are functions only of the 
wall-distance, the above expression approaches the following near-wall limit
\begin{equation}\label{ScRatio_NearWall}
R_{u'\phi'} \equiv \frac{ \overline{ u'_{2}\phi' } }{ \overline{ u'_{1}\phi'}} = 
\frac{R^{2}_{12}+R^{2}_{22}}{(R_{11}+R_{22})R_{12} } \rightarrow \frac{ R_{12} }{ R_{11} }\,,
\end{equation}
where subscripts 1, 2 and 3 are respectively the streamwise, wall-normal and spanwise directions.
Abe $\&$ Suga \cite{Abe2001} also observed that the correlation between $\phi'$ and $u'_{2}$ relatively 
increases with the decrease of $Pr$, reaching almost the same level as that between $\phi'$ and 
$u'_{1}$ for $Pr=0.025$. 
As a result, they recommended that an effective algebraic scalar-flux model should depend both
on linear and  quadratic forms of Reynolds stress, written as
\begin{equation}\label{abes_proposal}
\overline{ u'_{i}\phi' } = -\kappa \tau \bigg(  C_{\phi,1}\frac{R_{ij}}{\kappa} + C_{\phi,2}
\frac{ R_{ik}R_{kj} }{ \kappa^{2} }\bigg ) \Lambda_{j}\,,
\end{equation} 
where $C_{\phi,1}$ and $C_{\phi,2}$ are model coefficients. These
coefficients should be determined so that the non-linear term in equation~\eqref{abes_proposal} 
becomes dominant  in regions of high deformation rates, while the linear term becomes significant 
in the presence of weak strain rates, as well as in low Prandtl numbers. 
Hence, the present study aims to propose an algebraic model based on Younis general form
\eqref{general_representation} that involves the non-linear term, and validate its performance 
in different heated channel flows.

\section{PROPOSED FORMULATION}\label{proposed_formulation}

\begin{figure}[h!]
\flushleft
\subfloat[]
{\label{}
\includegraphics[width=0.49\textwidth]{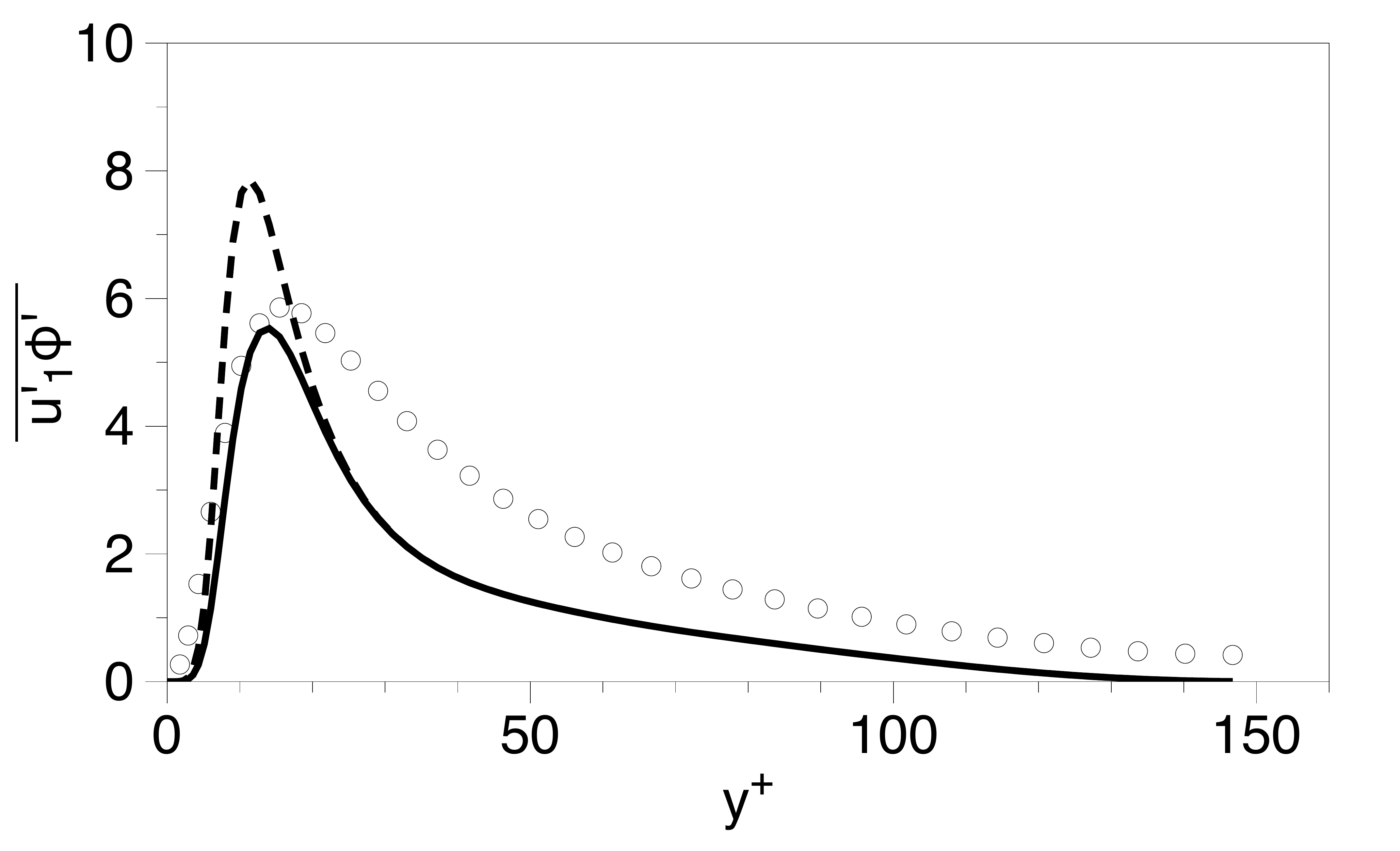}} 
\subfloat[]
{\label{}
\includegraphics[width=0.49\textwidth]{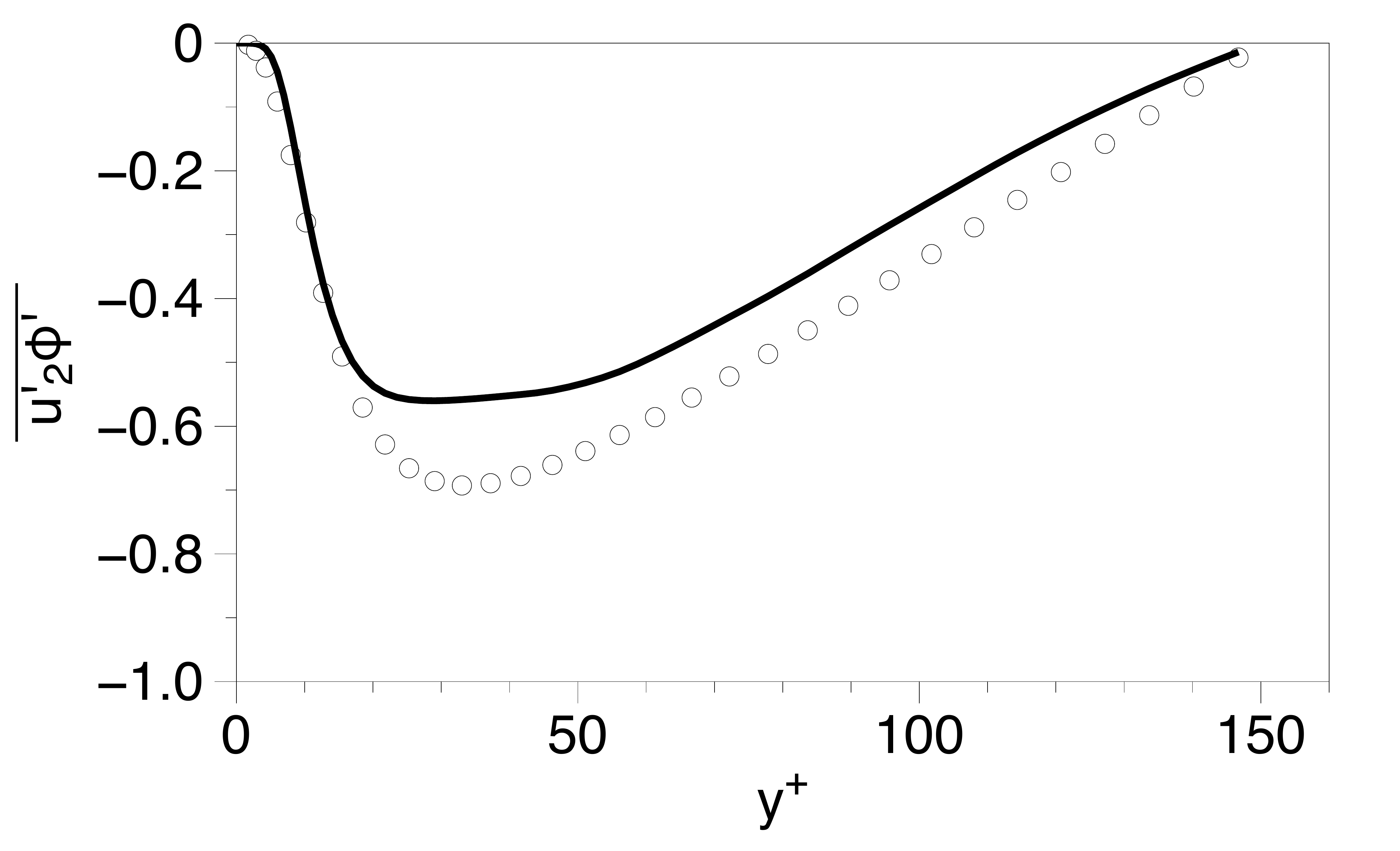}}\\
\flushleft
\subfloat[ ]
{\label{}
\includegraphics[width=0.49\textwidth]{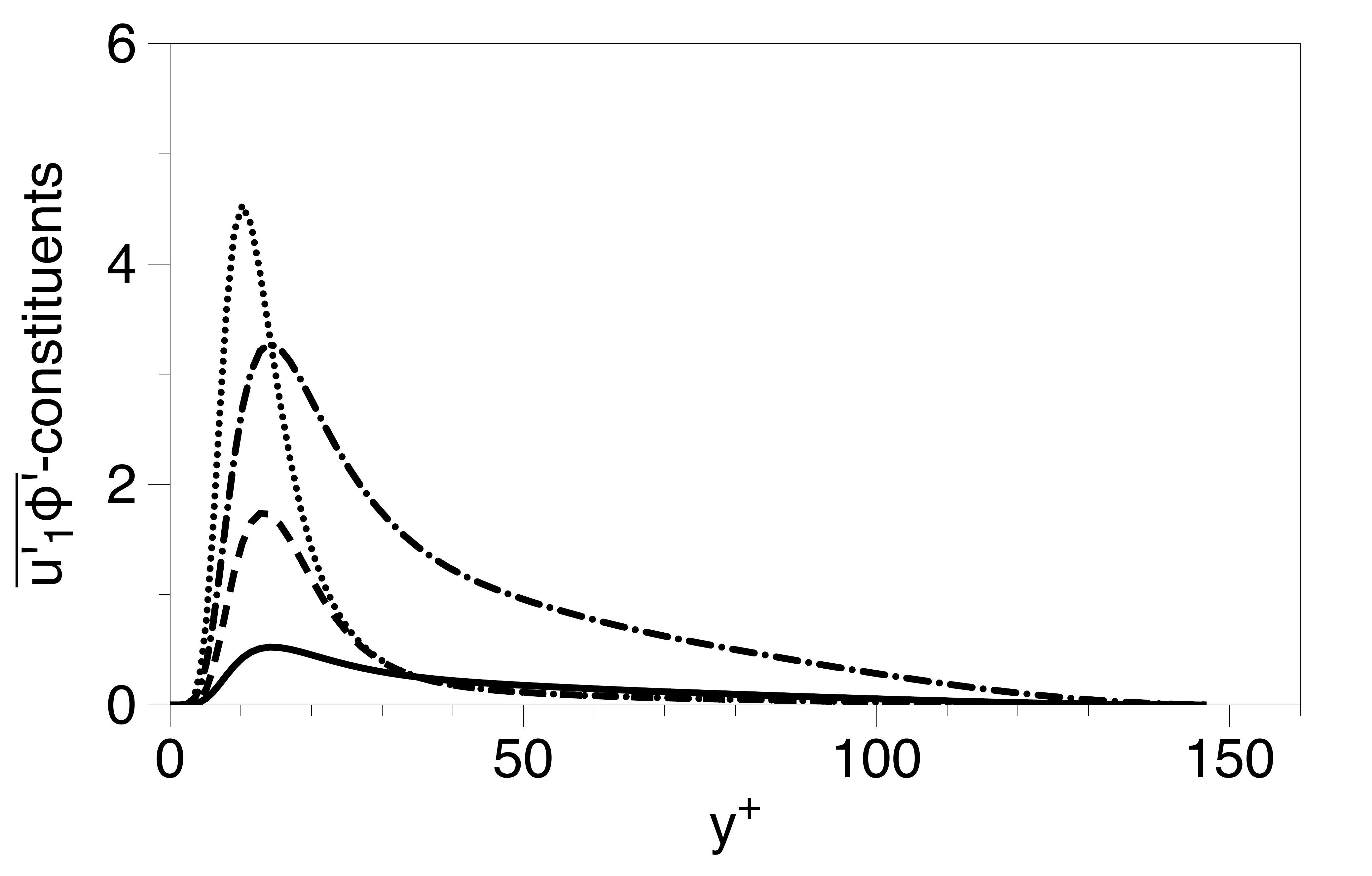}} 
\subfloat[  ]
{\label{}
\includegraphics[width=0.49\textwidth]{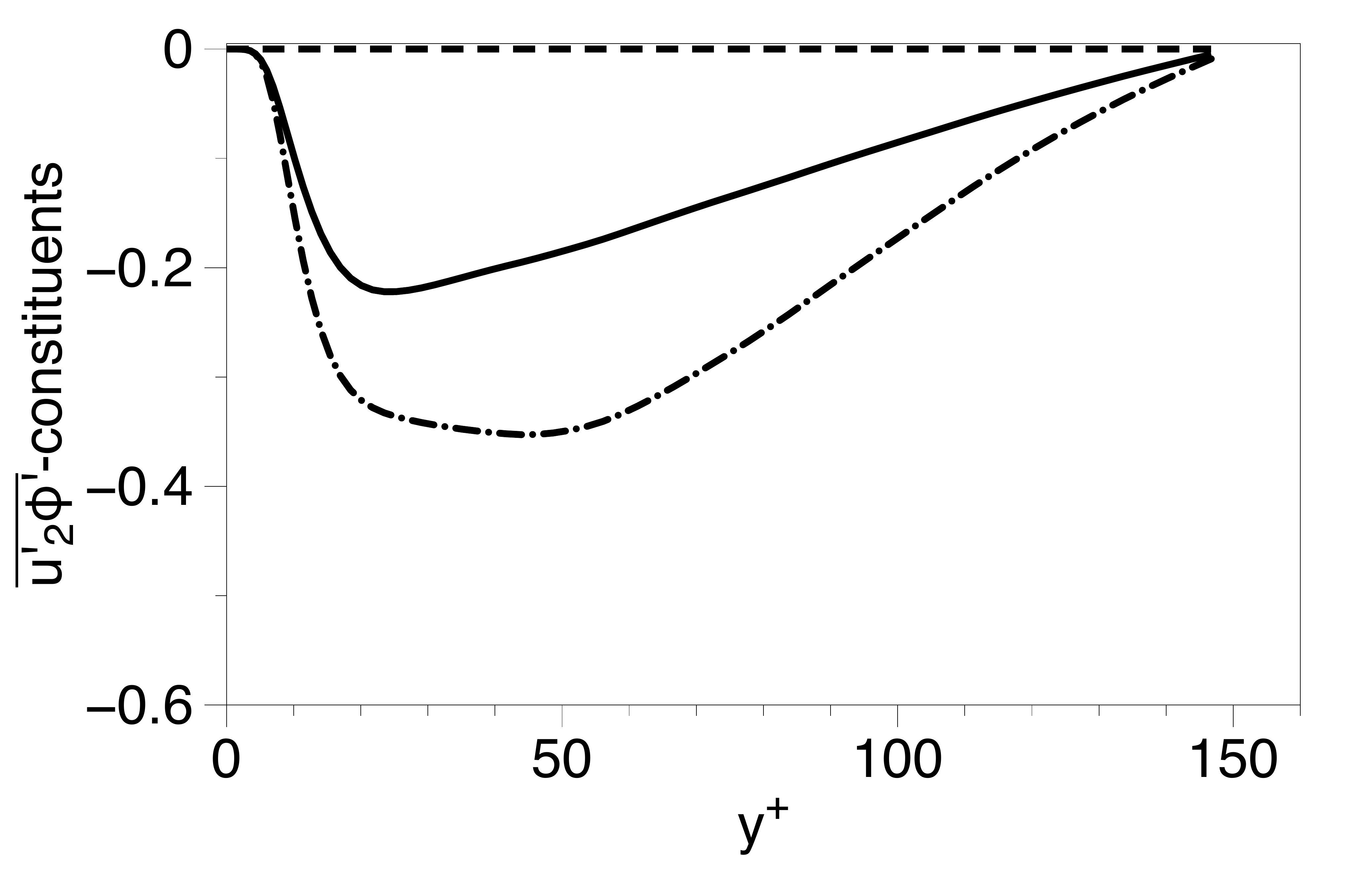}} 
\caption{ \textit{Top:} Predictions of the proposed model for the (a) streamwise and (b) 
normal scalar-flux components as a function of the distance from the wall $y^{+}$. Solid line ($\solid$) refers to the case where near-wall 
corrections are present \eqref{near_wall_corrections}, while dashed line 
($\dashed$)  refers to the uncorrected case \eqref{homogeneous_coefs}. 
Symbols denote the DNS results of Tomita et al. \cite{Tomita1993} for heated channel flow at 
$Re_{\tau}=150$ and $Pr=0.71$.
\textit{Bottom:} Balance between the individual terms appearing in the model equation 
\eqref{current_functional_form} for the (c) streamwise and (d) wall-normal scalar-flux components. 
Solid line ($\solid$) refers to the $C_{2}$-related term, dashed line ($\dashed$) refers to 
the corrected $C_{3}$-related term, dotted line ($\dotted$) refers to the uncorrected 
$C_{3}$-related term and dash-dotted ($\chndot$) line refers to the $C_{5}$-related term.
\label{fig:CALIBRATION_CASE}
} 
\end{figure}

In this section we introduce the proposed model, which
is essentially a combination between the multi-linear model of Younis \eqref{Younis_expression} and 
the functional form proposed by Abe $\&$ Suga \eqref{abes_proposal},
thus incorporating non-linear information regarding turbulence anisotropy.
Special attention is given to keep the model as simple as possible so that it can be easily implemented in
existing industrial codes. This is achieved by investigating the contribution of each term
appearing in the above equations in an attempt to propose a minimal combination of these 
terms that is able to provide reasonable predictions. 
We neglect the $C_{1}$-related term, since this term is known to provide 
erroneous predictions for the streamwise flux component. In addition, this choice can be 
partially justified 
if we assume that information regarding this term is already included in the 
isotropic part of the second term of equation~\eqref{Younis_expression}.
We have also excluded the term associated with $C_{4}$ for the following reasons:
Firstly, this term cannot induce the proper near-wall anisotropy levels of the scalar-flux vector, as 
given in equation~\eqref{ScRatio_NearWall}.
Secondly, this term involves products between the mean velocity gradients and the Reynolds
stress, thus being more complex than the $C_{3}$-related term, which also contains gradients
of the mean velocity field. Thirdly, its absence simplifies  the process of including wall-damping
corrections through the use of simple exponential functions, as will be shown in this Section. 
Consequently, the proposed closure takes the following compact form in a rotating frame of reference:
\begin{equation}\label{current_functional_form}
\overline{ u'_{i}\phi' } = C_{2}\frac{\kappa}{\epsilon} R_{ij}\Lambda_{j} 
+ C_{3} \frac{\kappa^{3}}{\epsilon^{2}}G^{T}_{ij}\Lambda_{j} 
+ \frac{ C_{5} }{ \epsilon } R_{ik}R_{kj} \Lambda_{j}\,,
\end{equation}
where we keep the same  indexing for the model coefficients as in equation~\eqref{Younis_expression}. 
Alternatively, the above equation essentially differs from equation~\eqref{abes_proposal} in involving the 
$C_{3}$-related term, which accounts for the products between mean scalar and mean velocity gradients. 
Model coefficients are determined through a two-step process. As a starting point, model 
constants ($C_{2}, C_{3}, C_{5}$) are determined through a linear regression fitting using data derived from the LES 
non-buoyant results of Kaltenbach \cite{kaltenbach1994} 
\begin{equation}\label{homogeneous_coefs}
C_{2} = -0.0848\,,\qquad C_{3} = 0.00496\,,\qquad C_{5} = -0.2942\,,
\end{equation} 
who considered  homogeneous shear flows under different orientations of the mean scalar 
gradient. The complete dataset used to determine model coefficients is given in the Appendix.
Next, we seek to modify the above coefficients to account for inhomogeneous effects,
particularly in the vicinity of wall-boundaries. As a result, we considered a 
fully-developed channel flow at friction Reynolds number $Re_{\tau}= \frac{ u_{\tau} \delta }{\nu}=150$, 
where $u_{\tau}$ is the friction velocity and $\delta$ is the half-channel height.  
The flow is driven by a mean pressure gradient, with uniform heat 
flux applied at both walls. The Prandtl number is set equal to 0.71, whereas the presented 
data is expressed in wall-units.
Figure~\ref{fig:CALIBRATION_CASE} shows model predictions for heat fluxes, obtained  by using 
DNS results of Tomita et al. \cite{Tomita1993} in a stationary frame.
For this case, only a streamwise component of the mean velocity field exists that 
varies along the normal component $x_{2}$. 
We observe that the proposed model overestimates the streamwise component, especially the 
near-wall peak magnitude, while reasonable predictions are provided for the normal component. 
For simplicity, we focus on improving only the streamwise component. 
A simple way to do that is by modifying $C_{3}$,  since the term associated with this 
coefficient does not contribute to the normal component, as indicated in equation~\eqref{current_functional_form}. 
Hence, we apply the following damping function to this coefficient to reduce its value 
close to the wall boundary,
\begin{equation}\label{near_wall_corrections}
C_{3} = 0.00496\,f_{C_{3}}\,,\qquad f_{C_{3}}= 1.0-\exp \bigg( -A\,\beta^{*}\,Pe^{\alpha^{*}} \bigg)\,,\qquad \alpha^{*}=-0.1\,,
\qquad \beta^{*}=30.0\,.
\end{equation}
\FloatBarrier
For $y^{+} < 15$, the  uncorrected $C_{3}$-related term plays a dominant role, with its maximum 
occurring at about $y^{+} = 10$, a location at which mean strain rate is also maximized (not shown here). 
The $C_{5}$-related term contributes the most for $y^{+}>15$, while the impact of 
$C_{2}$-related term becomes non-trivial in regions far away from the wall that are 
characterized by weak deformations. Applying the damping function on $C_{3}$ 
coefficient yields a reduction of the associated term, which remains important in the high-shear
region, while resulting in the quadratic term being dominant in the entire region outside the viscous sublayer ($y^{+}>5$).
Contrary, the $C_{3}$-related term does not contribute to the normal component, which is prevailed
by the non-linear term for $y^{+}>15$.  
Consequently, the near-wall behaviour of the model is driven by two mechanisms: one that involves 
products between mean velocity and scalar gradients, and a mechanism arising from the 
turbulence-turbulence interactions. 
Combining these two terms yields an alternative form for the proposed model
\begin{equation}\label{C2C3C5_effective}
\overline{ u'_{i}\phi' } = C_{2} \frac{\kappa}{\epsilon}R_{ij}\Lambda_{j} 
+ C_{3} \frac{\kappa^{3}}{\epsilon^{2}} G^{e}_{ij}\Lambda_{j}\,.
\end{equation} 
The above equation suggests that the non-linear interactions provide a gradient, in addition to
the actual mean gradient, thus yielding an effective gradient $G^{e}_{ij}$ 
\begin{equation}
G^{e}_{ij} = G^{T}_{ij} + \frac{ 4 C_{5} }{ C_{3} \tau }r_{ik}r_{kj}\,.
\end{equation}
This idea, called the ``effective-gradients" hypothesis, has been originally proposed by 
Kassinos \& Reynolds \cite{Kassinos1998} to construct a Reynolds-stress transport closure, and 
has been recently extended by Panagiotou \& Kassinos \cite{Panagiotou2016, Panagiotou2017, Panagiotou2020} 
for passive scalar transport. 
Thus, equation~\eqref{C2C3C5_effective} can be thought as an extension of Abe \& Suga proposal
\eqref{Abe_Suga_Proposal}, since it contains a linear term that becomes important in regions 
of weak deformation rate, while replacing the quadratic term by a term involving the effective
mean gradient that dominates the region of high and moderate deformations rates.

\section{MODEL ASSESSMENT}\label{MODEL_ASSES}  
  
\begin{figure}[h!]
\centering
\includegraphics[trim={8cm 0 7cm 0},clip,width=0.50\textwidth]{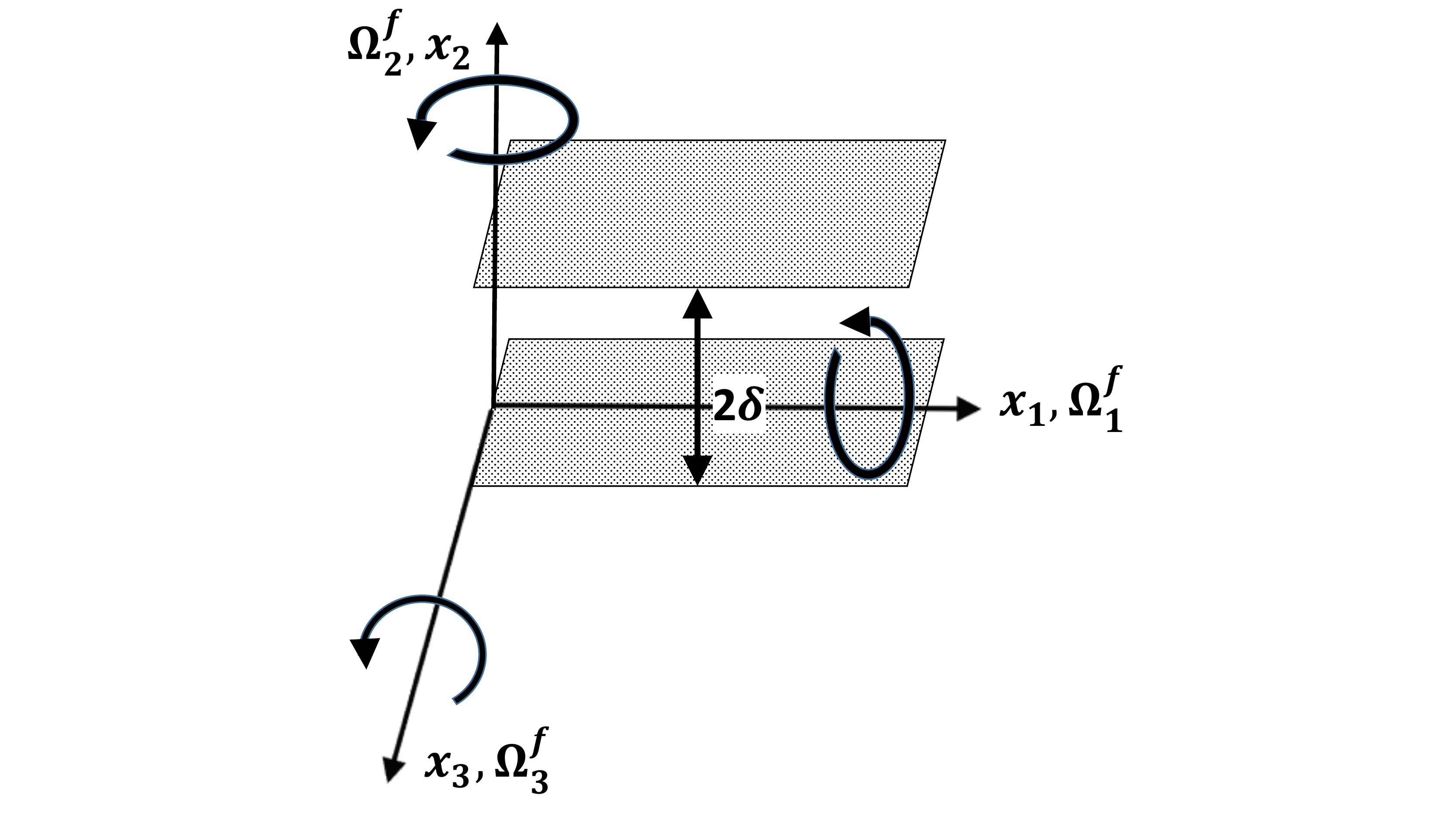}
\caption{ Flow configuration and coordinate system.
\label{fig:flow_config}
} 
\end{figure}

In this section we investigate the estimation ability of the proposed closure for heated 
channel flows under different boundary conditions and Reynolds numbers, in both stationary and rotating frames. For all cases considered, the Prandtl number 
equals to 0.71 and the flow variables are expressed in wall-units. In order to reduce 
uncertainties associated with the numerical solution of model equations, we import results 
from DNS into the proposed algebraic expression for the scalar-fluxes. 
To facilitate the discussion during the validation procedure, we have performed additional 
computations to account for the performance of Younis' model \eqref{Younis_expression}.  
For all cases considered, the mean flow varies only along the wall-normal direction $x_{2}$.
Under these conditions, and for a frame subjected to rotation around an arbitrary axis, the total
mean gradient $G^{T}_{ij}$ and the expressions for the flux components according to  
equation~\eqref{current_functional_form} become
\[
G^{T}_{ij} = 
 \begin{pmatrix}
     0              & G_{12}-\Omega^{f}_{3}    &   \Omega^{f}_{2}  \\
   \Omega^{f}_{3}   & 0                        &  -\Omega^{f}_{1}  \\
  -\Omega^{f}_{2}   & G_{32} + \Omega^{f}_{1}  &   0               \\
 \end{pmatrix}
 ,
\]
and
\begin{subequations}\label{complete_model_equations}
\begin{align}
\overline{ u'_{1}\phi' } & = \bigg \{ C_{2}\frac{\kappa}{\epsilon} R_{12} + C_{3}\frac{\kappa^{3}}{\epsilon^{2}}
(G_{12}-\Omega^{f}_{3}) + \frac{C_{5}}{\epsilon}\bigg( R_{12}(R_{11}+R_{22}) + R_{13}R_{23} \bigg) \bigg \}
\Lambda_{2}\,,  \label{u1phi_eq} \\
\overline{ u'_{2}\phi' } & = \bigg \{ C_{2}\frac{\kappa}{\epsilon} R_{22} 
+ \frac{C_{5}}{\epsilon}\bigg(  R^{2}_{12} + R^{2}_{22} + R^{2}_{23} \bigg) \bigg \} \Lambda_{2}\,, \label{u2phi_explicit_eq} 
\end{align}
\end{subequations}
respectively.
The flow geometry and coordinate system are shown in figure~\ref{fig:flow_config}.
\FloatBarrier

\subsection{Heated Couette flows}

\begin{figure}[h!]
\flushleft
\subfloat[]
{\label{}
\includegraphics[width=0.49\textwidth]{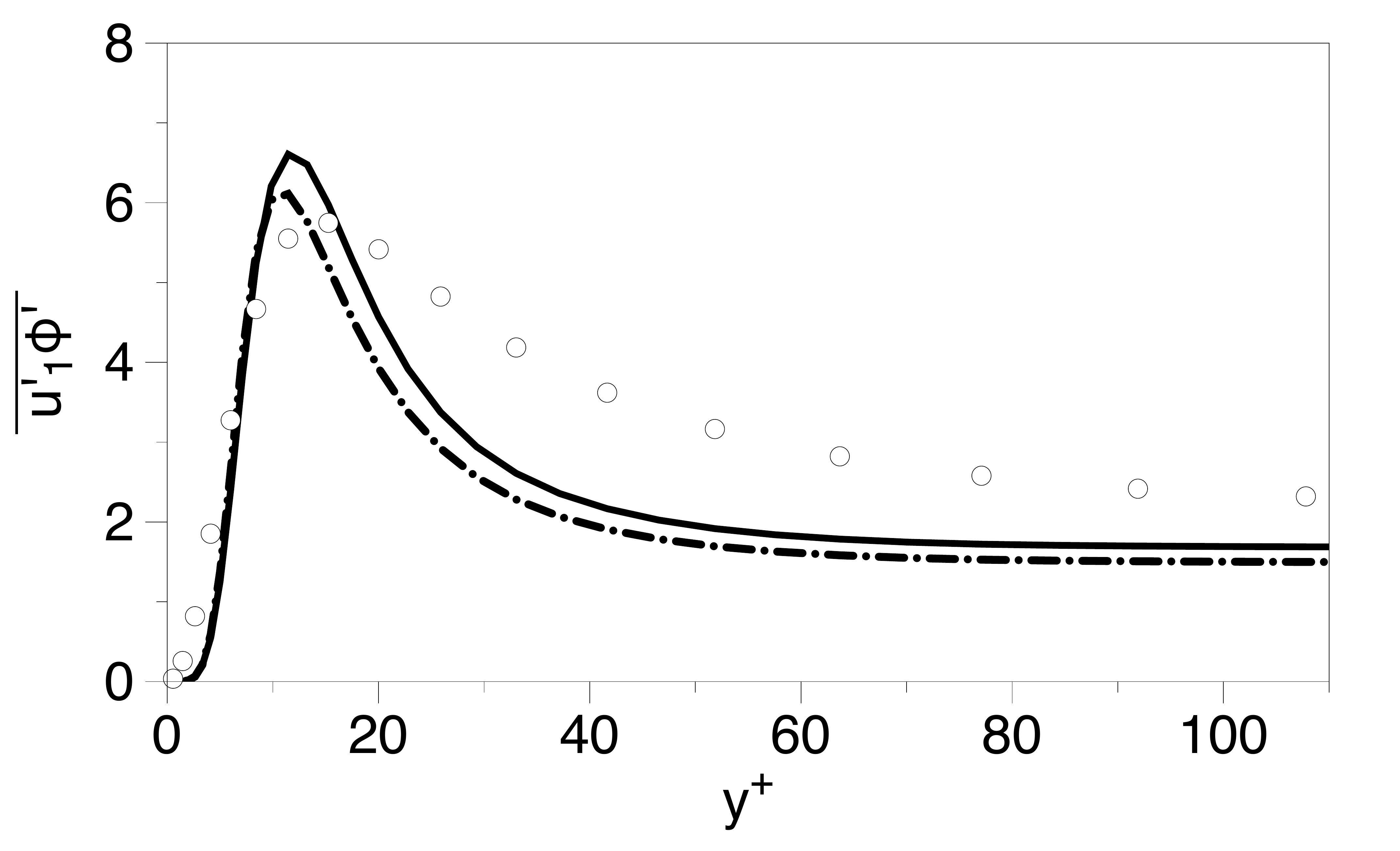}} 
\subfloat[]
{\label{}
\includegraphics[width=0.49\textwidth]{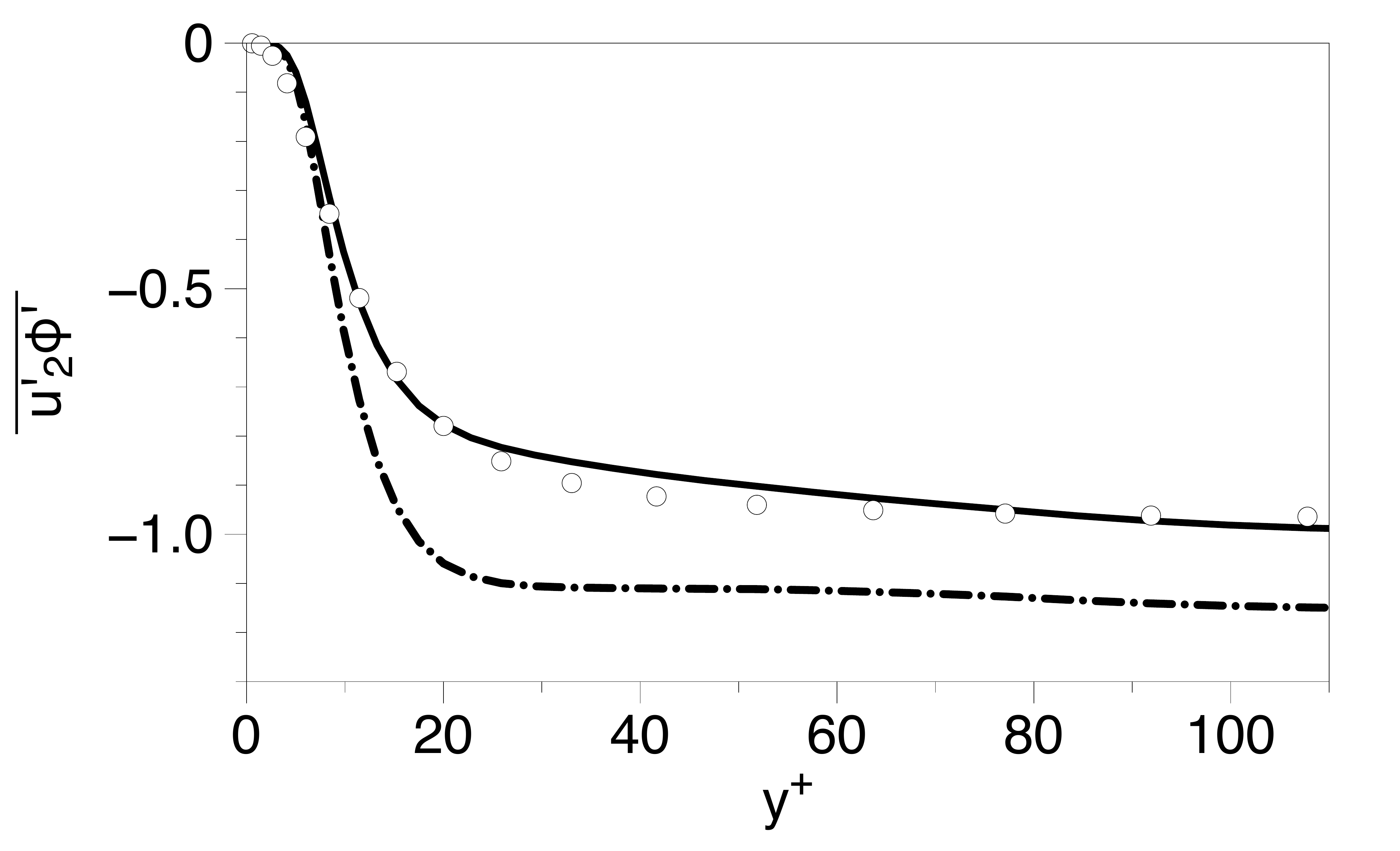}}
\caption{ Model predictions for the (a) streamwise and (b) normal scalar-flux components  
for a heated Couette flow at $Re_{w}=8600$. Solid line ($\solid$) denotes the proposed model
and dash-dotted line ($\chndot$) denotes Younis model. Comparison is made with DNS results
of Kawamura et al. \cite{Kawamura2000}.
\label{fig:COUETTE_ScF_8600}
} 
\end{figure}

\begin{figure}[h!]
\flushleft
\subfloat[]
{\label{}
\includegraphics[width=0.49\textwidth]{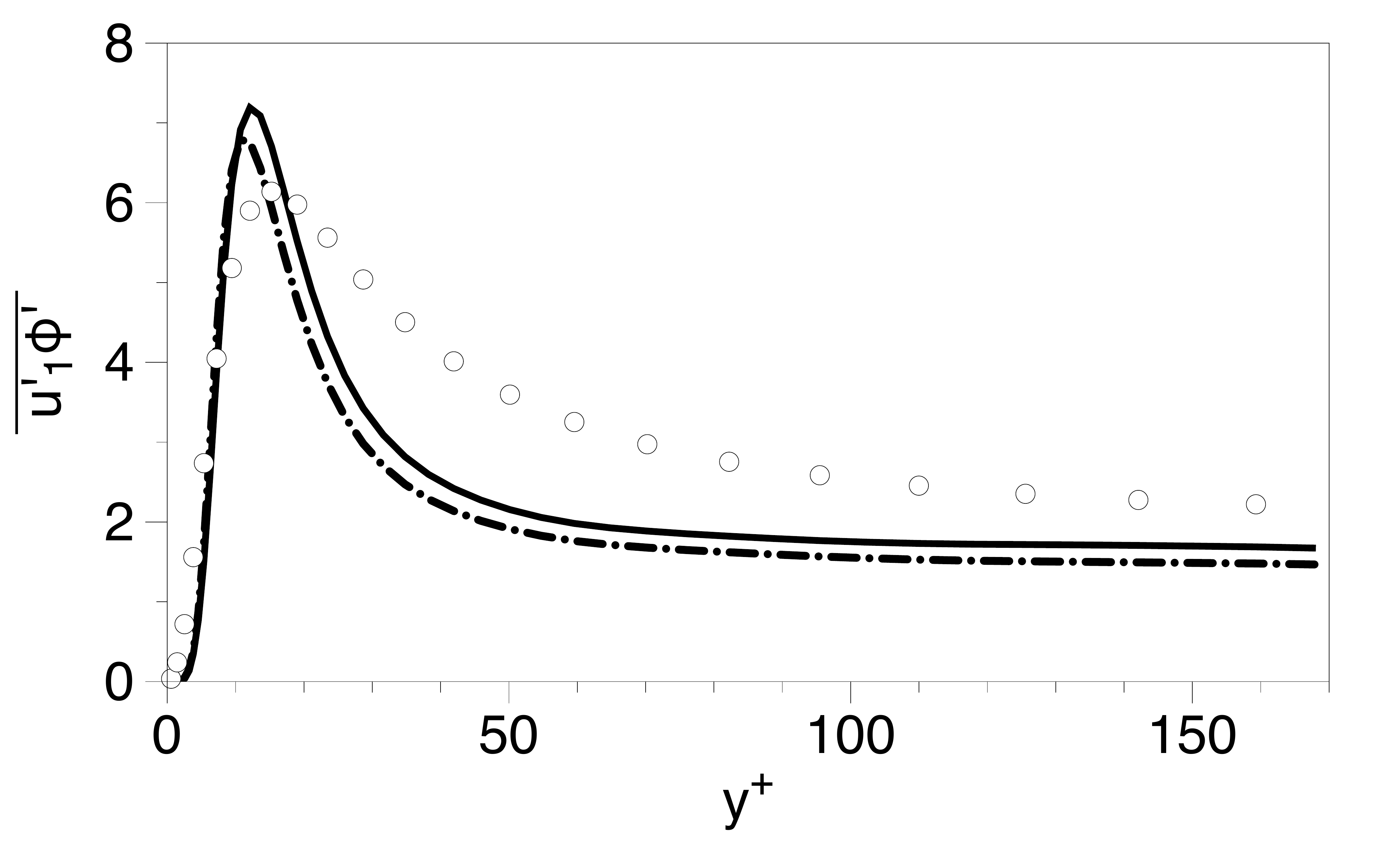}} 
\subfloat[]
{\label{}
\includegraphics[width=0.49\textwidth]{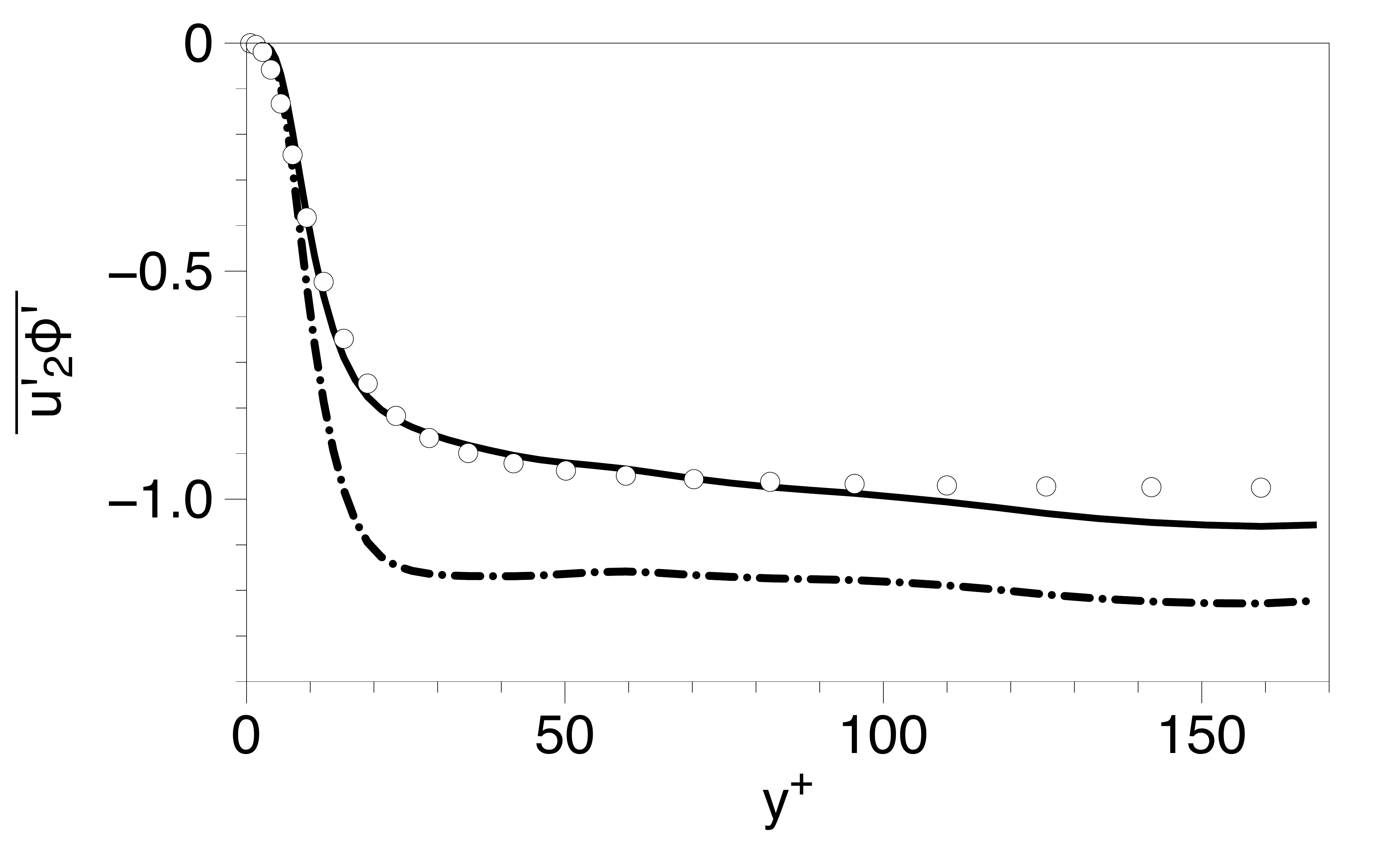}}\\
\caption{ As in Figure~\ref{fig:COUETTE_ScF_8600}  but for $Re_{w}=12800$.
\label{fig:COUETTE_ScF_12800}
} 
\end{figure}

Initially, we evaluate the model performance in a fully-developed Couette flow with the top 
wall moving at constant speed $u_{w}$ relative to the bottom wall. As a result, the mean 
flow is driven by the shear stress due to the relative movement between top and bottom walls,
while the temperature difference between the heated top wall and the cooled bottom wall is 
kept constant. 
We consider two cases at different Reynolds numbers, particularly $Re_{w}=\frac{2\,u_{w}\,\delta}{\nu}=8600$ and 12800, 
for which detailed DNS data are provided by Kawamura et al. \cite{Kawamura2000}. 
Figures~\ref{fig:COUETTE_ScF_8600} and \ref{fig:COUETTE_ScF_12800} show a comparison between
the two models (proposed and Younis) for the scalar-flux components with the corresponding 
DNS results for the 
two cases. Regarding the streamwise component, both models perform well near the wall boundary, 
being able to capture both the 
magnitude and the location of the near-wall peak, while both closures predict faster reduction 
of the shape profile with respect to the DNS results while moving away from the wall. 
The proposed model agrees well with the DNS data for the wall-normal component, especially 
at the wall region and close to the channel center, while Younis' model underestimates the 
wall-normal component outside the viscous sublayer ($y^{+}>10$). 
\FloatBarrier

\FloatBarrier

\subsection{Heated Poiseuille flows}

\begin{figure}[h!]
\flushleft
\subfloat[]
{\label{}
\includegraphics[width=0.49\textwidth]{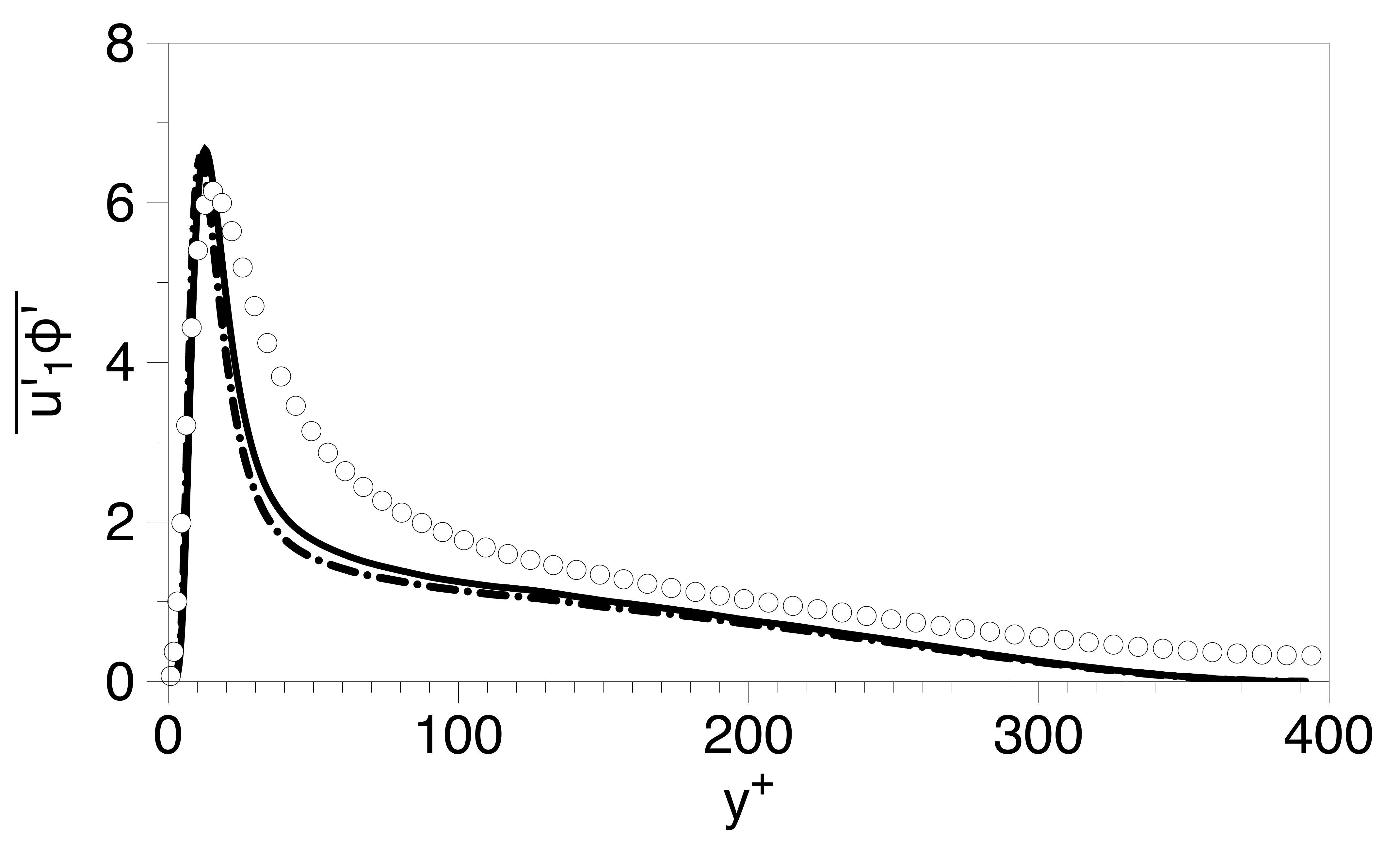}} 
\subfloat[]
{\label{}
\includegraphics[width=0.49\textwidth]{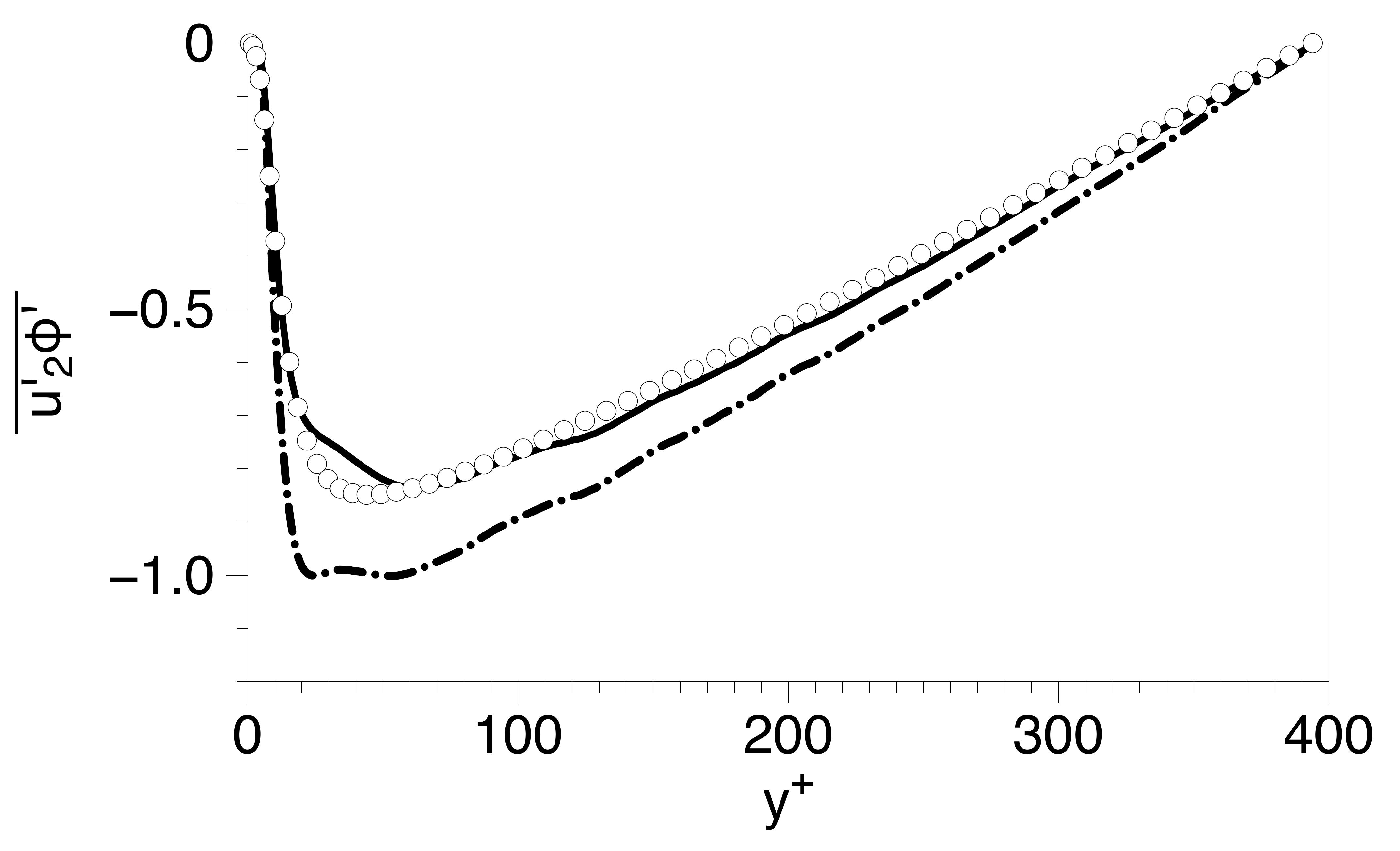}}\\
\subfloat[]
{\label{}
\includegraphics[width=0.49\textwidth]{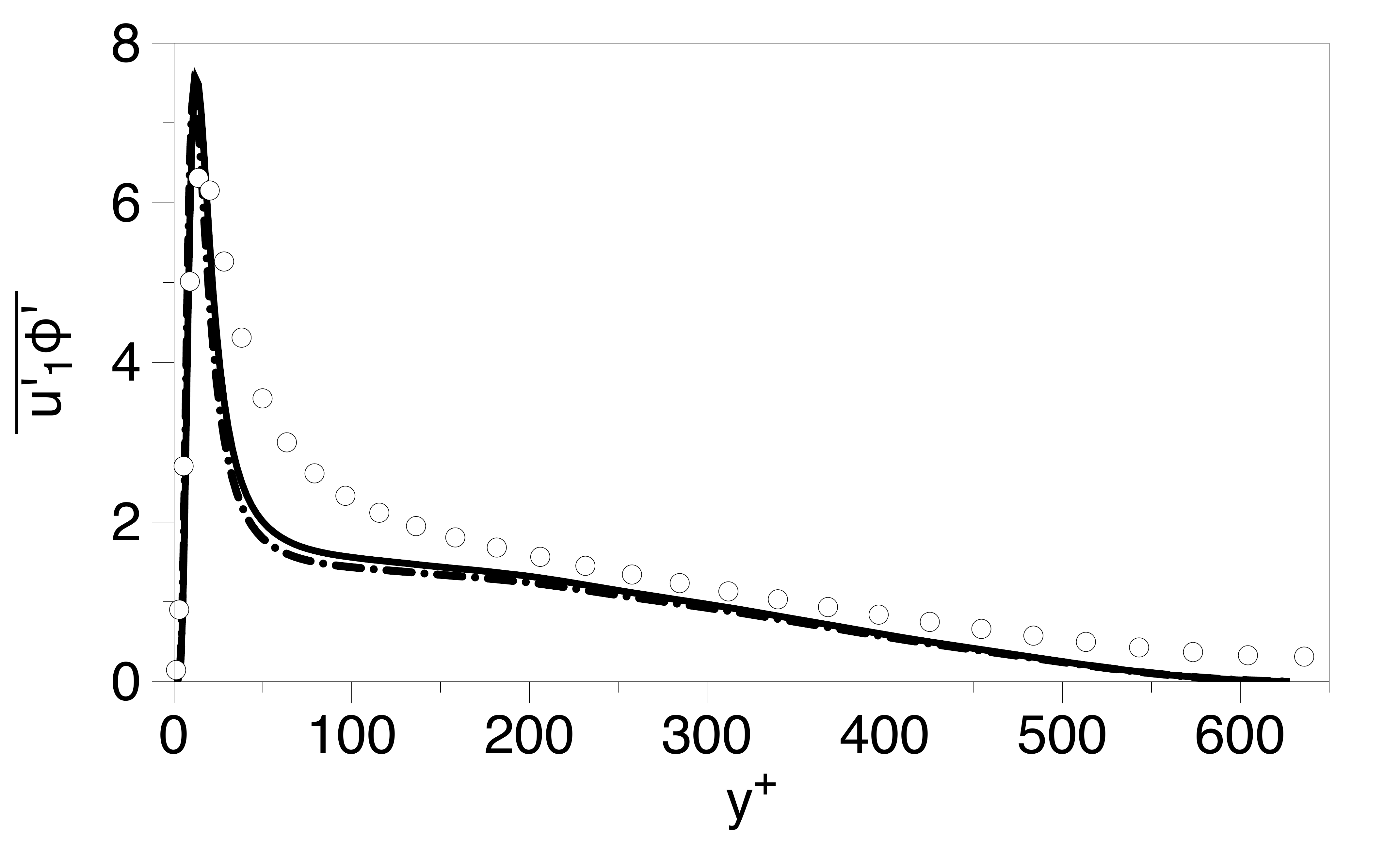}} 
\subfloat[]
{\label{}
\includegraphics[width=0.49\textwidth]{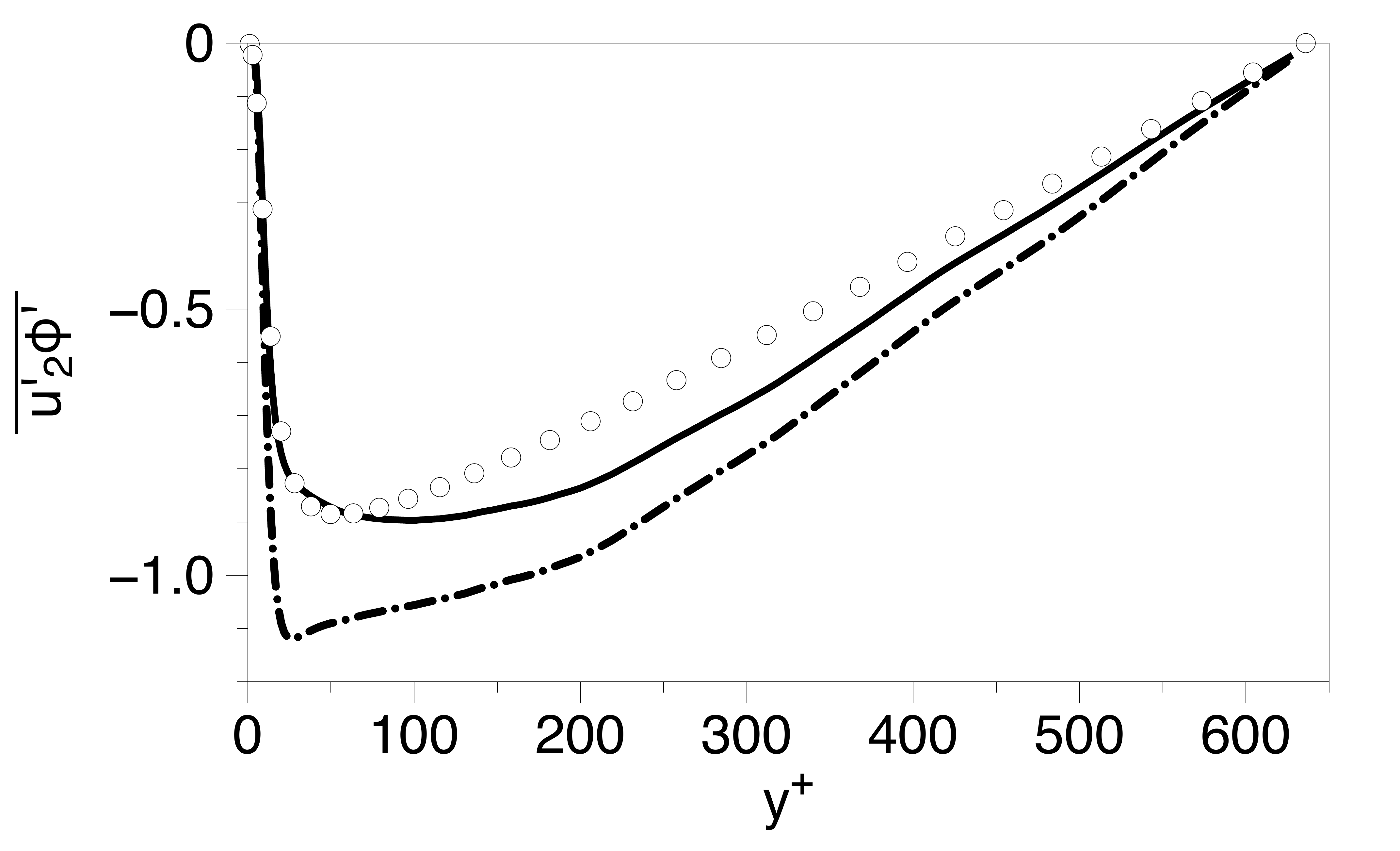}}
\caption{ Model predictions for the scalar-flux components for a heated 
channel flow: $Re_{\tau}=395$ (a-b), $Re_{\tau}=640$ (c-d). Solid line ($\solid$) denotes 
the proposed model and dash-dotted line ($\chndot$) denotes Younis' model. Symbols denote the
DNS results. 
\label{fig:POISEUILLE_ScF_395_640}
} 
\end{figure}

Here, we consider the case of a heated channel flow under fully-developed conditions at 
two different Reynolds numbers, particularly $Re_{\tau}=395$ and 640. 
The non-slip boundary condition is adopted in the wall-normal direction (i.e. at the top 
and bottom walls) while the thermal boundary condition is uniform heat-flux on both 
walls.  The quality of the predictions is compared with the corresponding DNS results 
\cite{Abe2001, Abe2004}. Figure~\ref{fig:POISEUILLE_ScF_395_640} compares the turbulent 
scalar-fluxes as obtained from the present model and the Younis model for both cases. We observe 
that both models provide similar predictions for the streamwise component, achieving good 
agreement with the DNS data in both the inner and outer region of the channel. For the high 
$Re$ case though, both models tend to overpredict the magnitude of the peak value, thus 
showing a mild Reynolds dependence. 
Looking into the balance between the terms appearing in 
equation~\eqref{u1phi_eq} (not shown here) revealed that this tendency is mainly attributed 
to the $C_{3}$-related term, since this term exhibits the highest sensitivity to the Reynolds number.
Contrary to algebraic closures, DNS predict non-zero values for the streamwise component at 
the channel centerline. This happens because the turbulent scalar field is being transported 
by the mean flow or the turbulence. These transport processes constitute a non-local mechanism
that cannot be captured by a rational algebraic closure \cite{Younis2005}.
For the low $Re$ case, the proposed model achieves a considerably better agreement than Younis'
model regarding the normal flux, being able to quickly adjust to the slope's sign change that occurs in the inner region.
As expected, both models tend to overestimate the magnitude of the flux for $Re_{\tau}=640$, 
with the proposed model still being able to produce satisfactory results. 
\FloatBarrier

\begin{figure}[h!]
\flushleft
\subfloat[]
{\label{}
\includegraphics[width=0.49\textwidth]{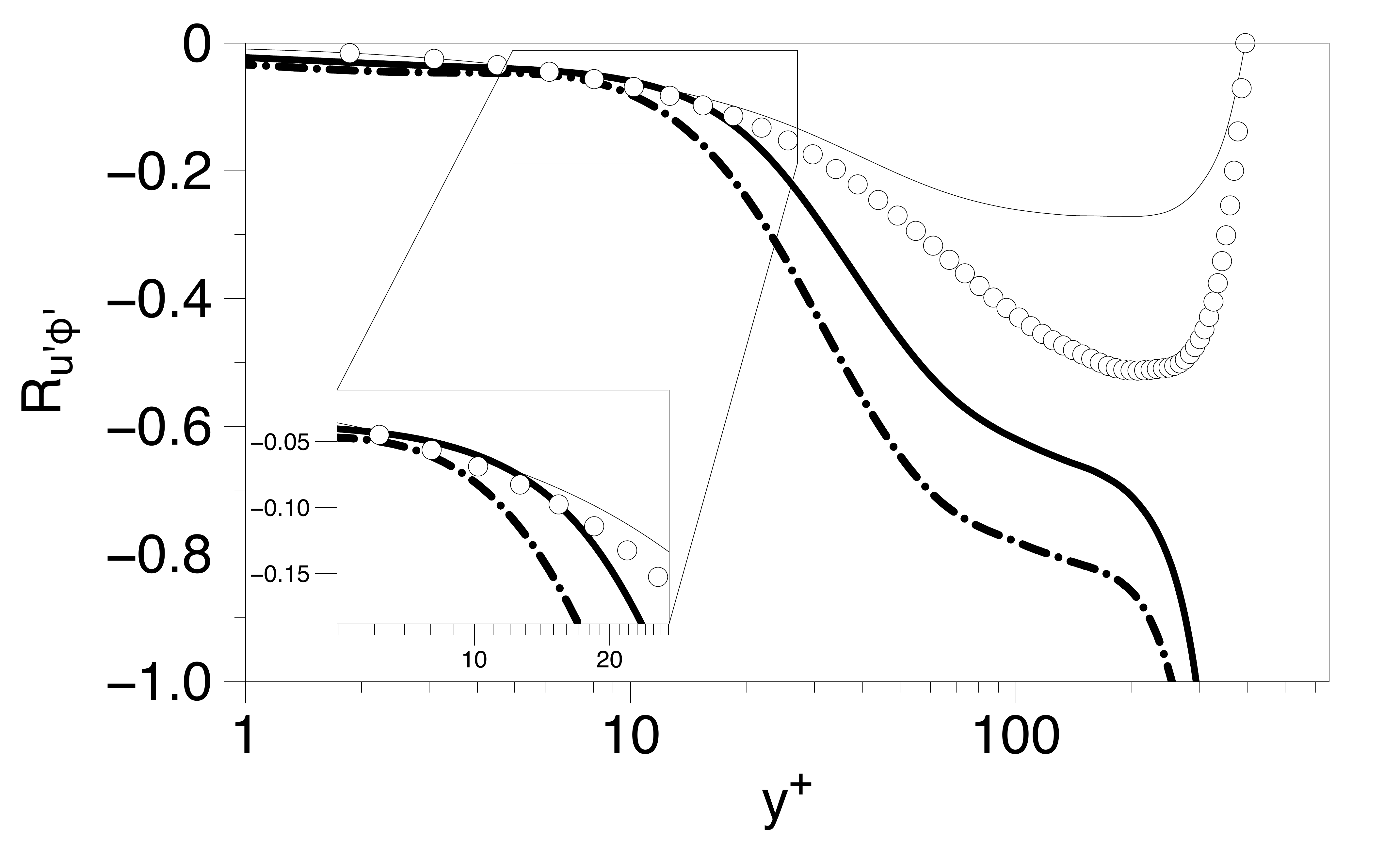}} 
\subfloat[]
{\label{}
\includegraphics[width=0.49\textwidth]{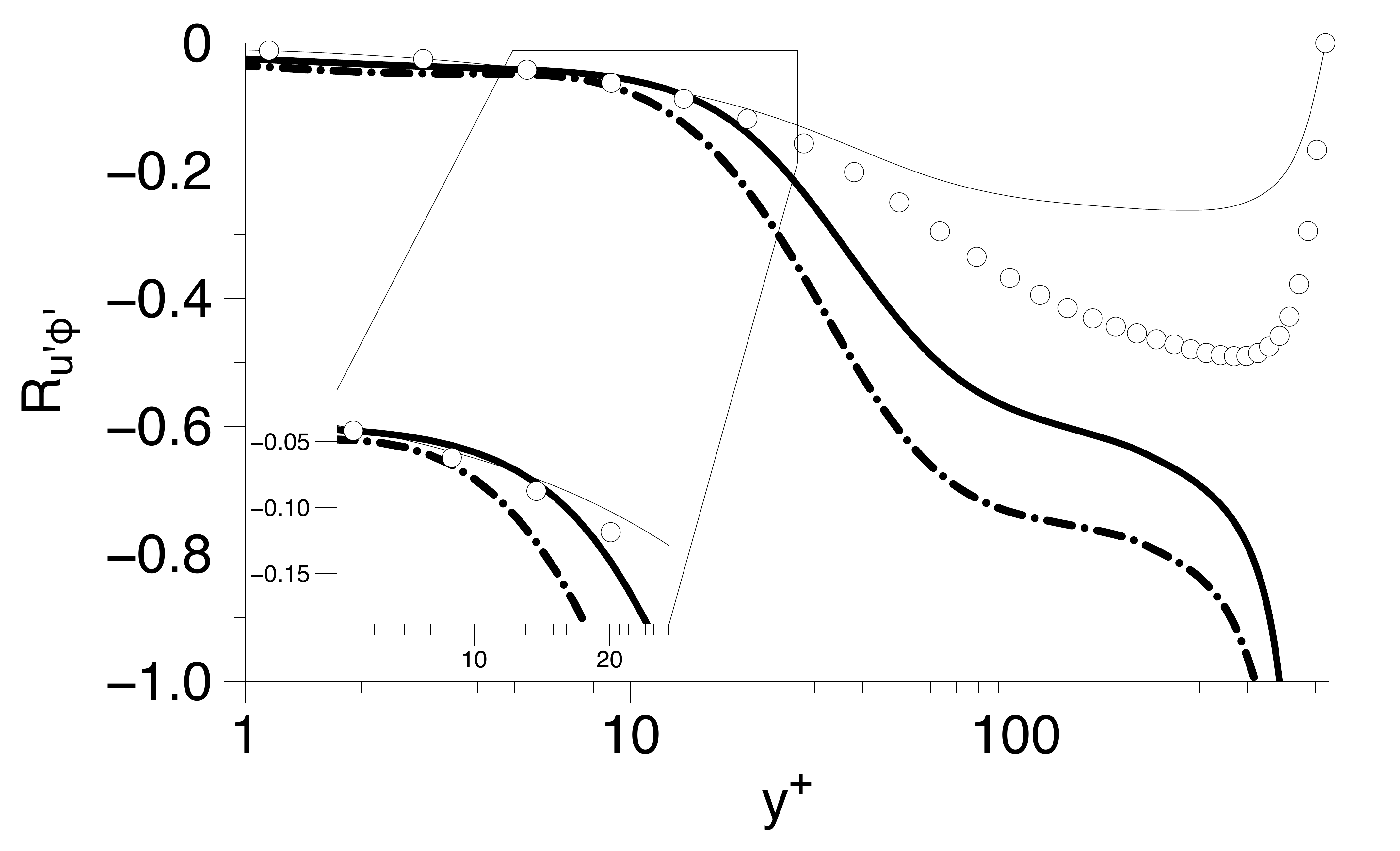}}\\
\caption{ Model predictions for the scalar ratio $R_{u'\phi'}$ for a heated Poiseuille flow 
at (a) $Re_{\tau}=395$  and (b) $Re_{\tau}=640$. Comparison is made with DNS predictions
for the stress ratio $R_{12}/R_{11}$ (thin lines) and the scalar ratio (symbols). Solid line 
($\solid$) denotes the proposed model and dash-dotted line ($\chndot$) denotes Younis' model.
Zoomed view refers to the near-wall region. 
\label{fig:POISEUILLE_ScR}
} 
\end{figure}

Next, we further investigate the near-wall performance of the proposed closure. A useful parameter
for that purpose is the scalar-flux ratio $R_{u'\phi'}$, defined in equation~\eqref{ScRatio_NearWall}.
Figure~\ref{fig:POISEUILLE_ScR} shows a comparison of the scalar-flux ratio predictions of both
models (proposed and Younis) with the
corresponding DNS data, for which this ratio varies as $R_{12}/R_{11}$ in the 
near-wall region under high shear strain, as already mentioned in Section~\ref{NLTERM_INCLUSION}.
The proposed model achieves better agreement with the DNS results, being able to capture 
the near-wall limit up to $y^{+}=20$. This is attributed to the dominant role that $C_{5}$-related
term plays in the buffer layer for both flux-components, as already shown in figure~\ref{fig:CALIBRATION_CASE},
which exhibits the proper near-wall physical behaviour (see discussion in Section~\ref{NLTERM_INCLUSION}).
In contrast, Younis' model fails to capture this limiting behaviour for $y^{+}>10$, since
no term appearing in its model equation \eqref{Younis_expression} captures the correct turbulent anisotropy.
\FloatBarrier

\subsection{Heated channel flow subjected to streamwise rotation}

\begin{figure}[h!]
\flushleft
\subfloat[]
{\label{}
\includegraphics[width=0.49\textwidth]{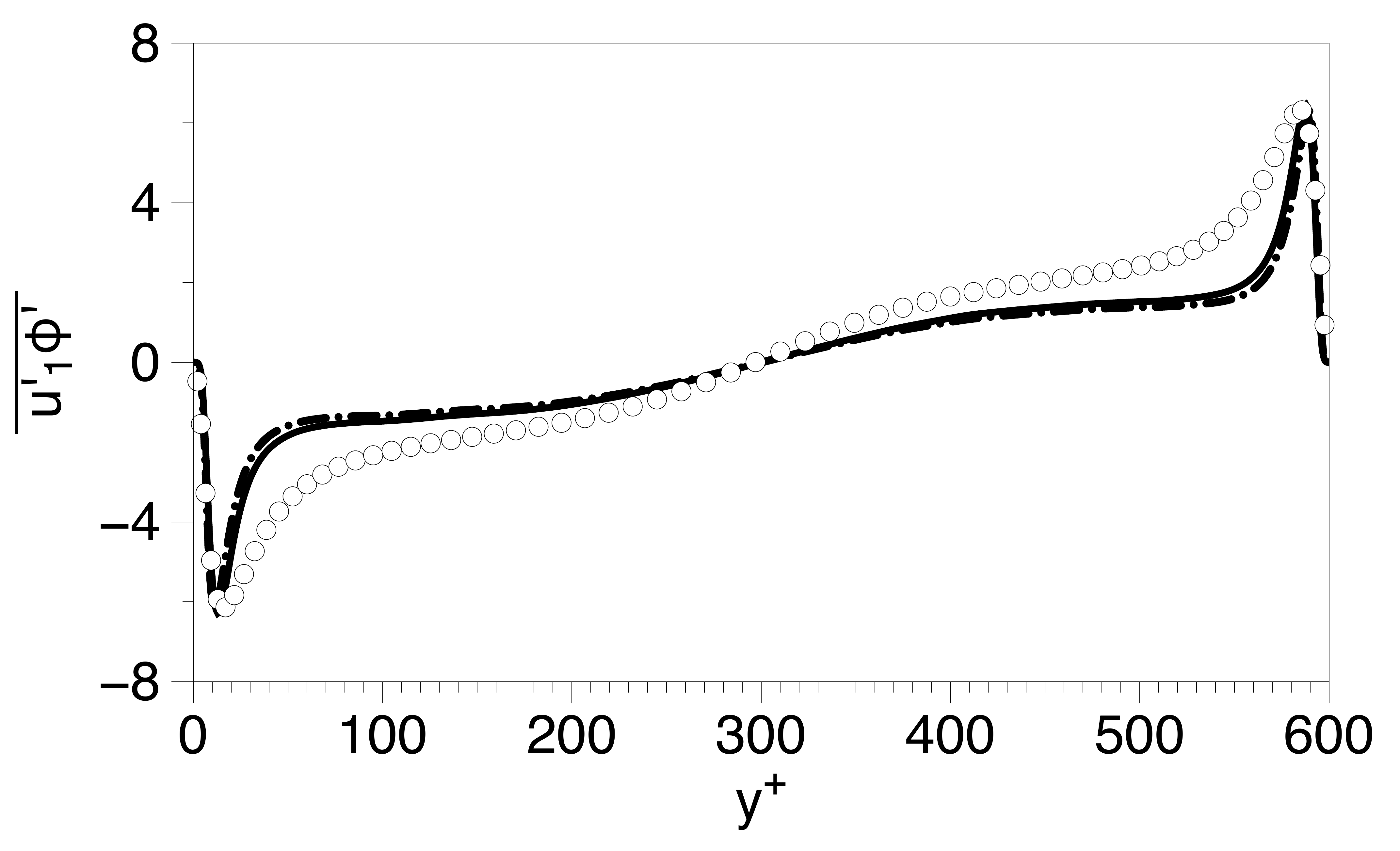}} 
\subfloat[]
{\label{}
\includegraphics[width=0.49\textwidth]{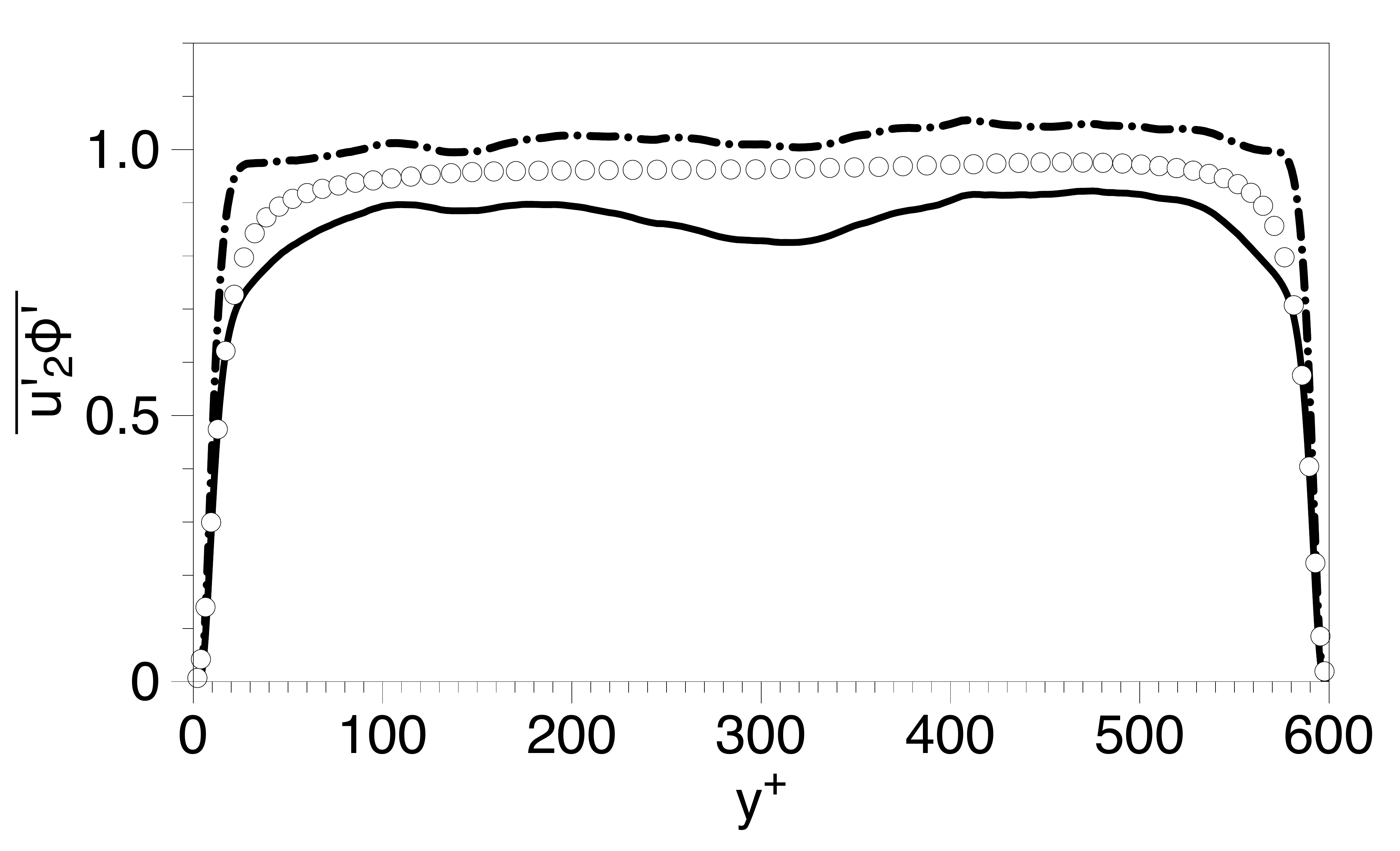}}\\
\caption{ Model predictions for the (a) streamwise and (b) wall-normal scalar flux components
for a heated channel flow at $Re_{\tau^{*}}=300$ that rotates around the streamwise axis at
$Ro_{\tau}=2.5$. Solid line ($\solid$) denotes the proposed model and dash-dotted line 
($\chndot$) denotes Younis' model. Symbols denote the DNS results of Wu \& Kasagi \cite{Wu2004}.
\label{fig:RE_300_Ro25_ST}
} 
\end{figure}

\begin{figure}[h!]
\flushleft
\subfloat[]
{\label{}
\includegraphics[width=0.49\textwidth]{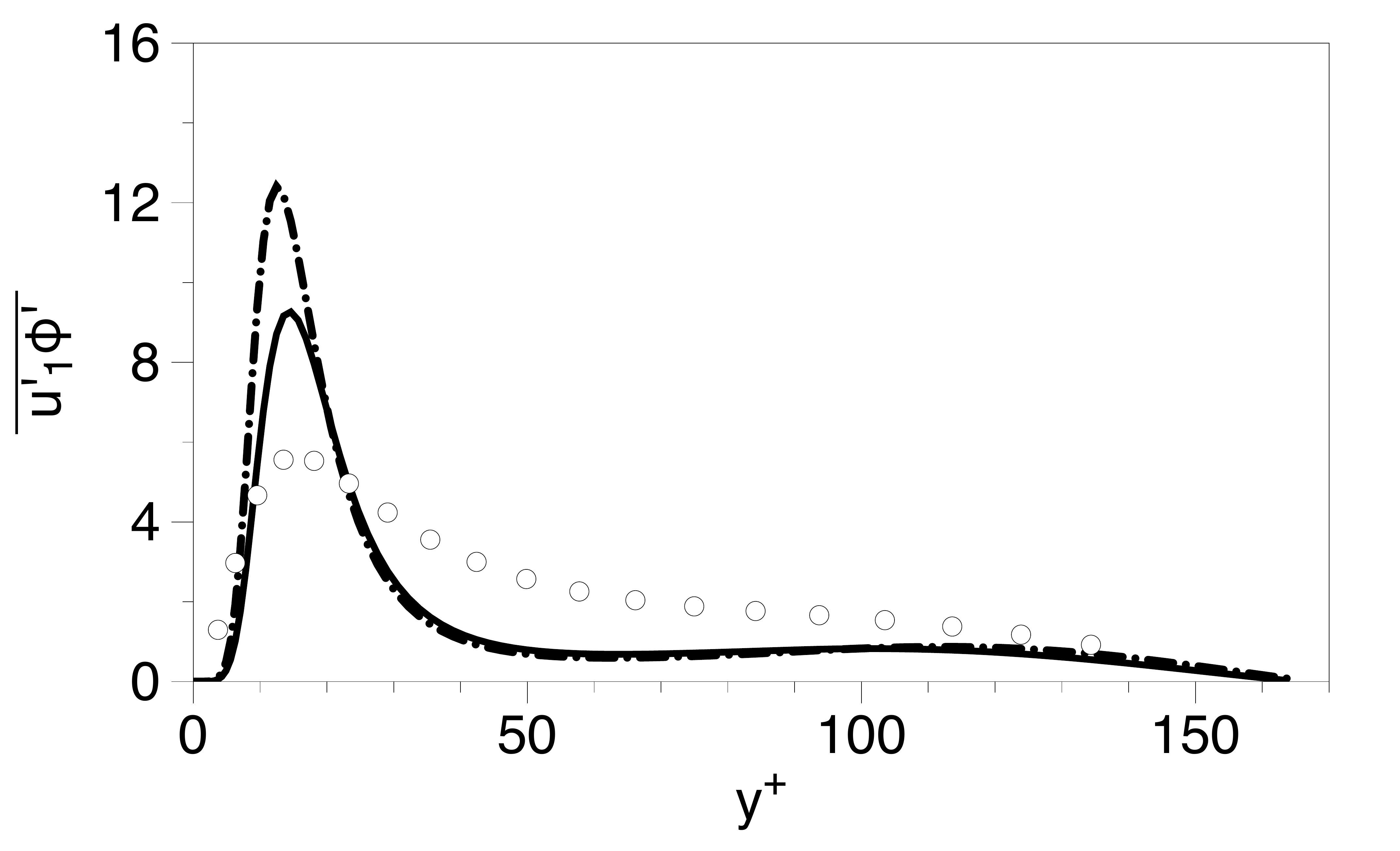}} 
\subfloat[]
{\label{}
\includegraphics[width=0.49\textwidth]{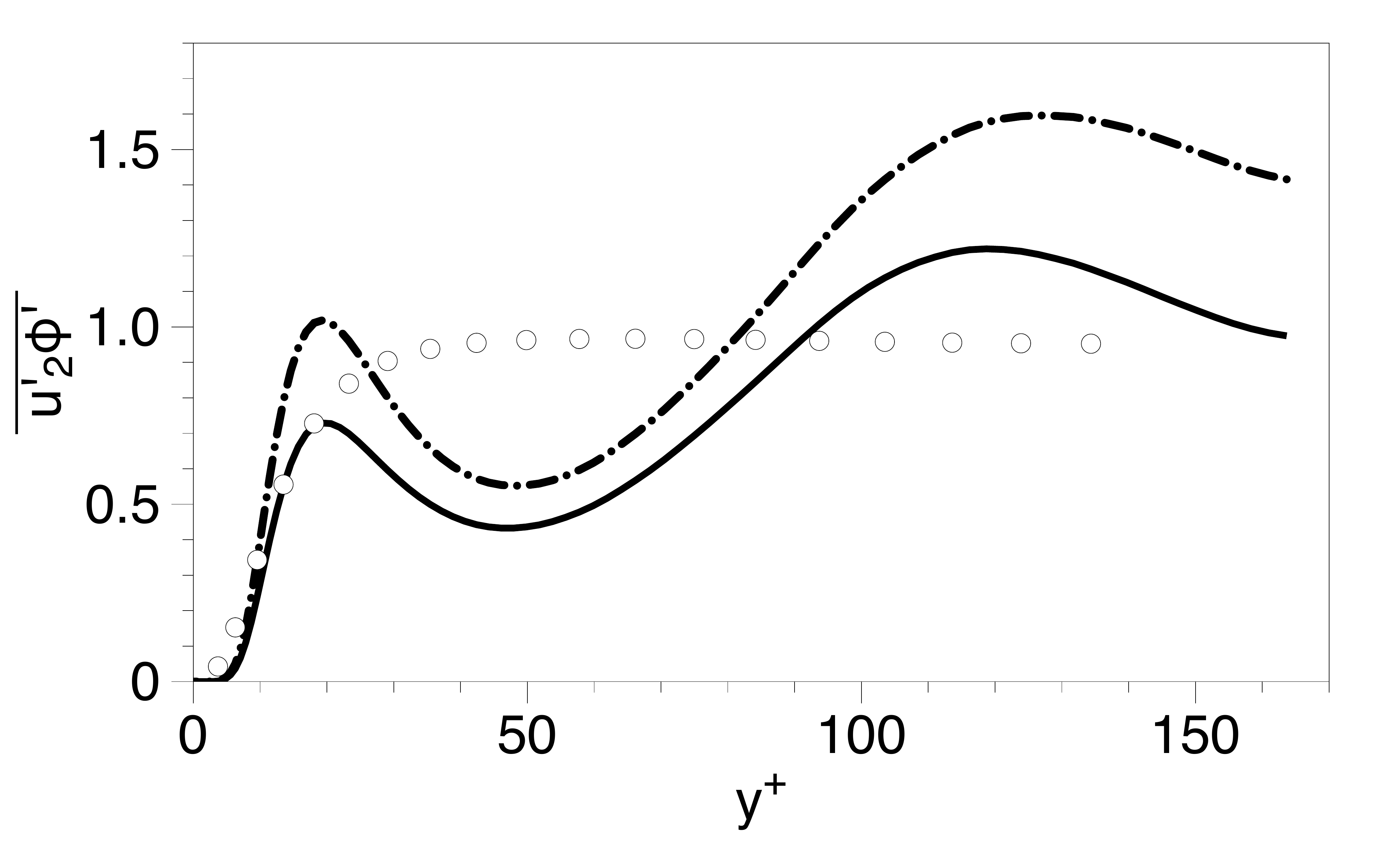}}\\
\caption{ Model predictions for the (a) streamwise and (b) wall-normal scalar-flux components
for a heated channel flow at $Re_{\tau^{*}}=163$ that rotates around the streamwise axis at
$Ro_{\tau}=15$. Solid line ($\solid$) denotes the proposed model and dash-dotted line 
($\chndot$) denotes Younis' model. Symbols denote the DNS results of  
El-Samni \& Kasagi \cite{ElSamni2000}.
\label{fig:RE_163_Ro15_ST}
} 
\end{figure}

The first considered case involving Coriolis effects is that of a flow in a plane channel that 
rotates about its streamwise axis. This scenario is more complex than the stationary one, 
since it induces a non-trivial spanwise mean velocity component, thus activating additional  
Reynolds stress elements. 
The walls are assumed to be kept at different, but constant temperatures without fluctuations,
with the upper wall cooled  and the lower wall heated.
Comparison is made with DNS studies that differ substantially in the relative strength of
the imposed rotation. 
The first study corresponds to the DNS results of Wu \& Kasagi \cite{Wu2004} for Rossby and 
Reynolds numbers $Ro_{\tau}= \frac{ 2 \Omega^{f}_{1}\delta}{u^{*}_{\tau}}=2.5$ and 
$Re_{\tau^{*}}= \frac{ u^{*}_{\tau} \delta }{\nu}= 300$ respectively, where $u^{*}_{\tau}$ is the friction
velocity calculated from the wall shear stress averaged on the two walls.
Figure~\ref{fig:RE_300_Ro25_ST} shows results for the scalar-flux components. For the 
streamwise component, both models accurately capture the location and magnitude of the peak 
value. However, similar to the stationary cases, their predictions drop much faster than the DNS 
results while approaching the channel's centerline. The present closure provides reasonable
estimations for the normal component, being able to capture the near-wall anisotropy, while
it mildly underestimates the level of this quantity while moving away from the wall. This 
happens possibly due to the fact that the effects of streamwise rotation on the normal 
component are explicitly absent, as indicated in equation~\eqref{u2phi_explicit_eq}. The
opposite is true for Younis' formulation, for which this dependence enters via the $C_{4}$-related
term, as implied by equation~\eqref{Younis_expression}.
\FloatBarrier
The second case study considered is by El-Samni \& Kasagi \cite{ElSamni2000} for $Re_{\tau^{*}} =163$ 
in the presence of a much stronger rotation rate, corresponding to $Ro_{\tau}=15$, whereas the 
thermal boundary conditions are the same as in the low-$Ro_{\tau}$ case. 
Figure~\ref{fig:RE_163_Ro15_ST} shows profiles of the scalar-flux components, showing large
discrepancies between DNS and models. Both models
significantly overpredict the maximum value of the streamwise component. Regarding the proposed 
closure, this is partly 
associated with the presence of non-trivial contribution from the product $R_{12}\,R_{23}$ that appears in the 
$C_{5}$-related term of the model equation \eqref{u1phi_eq}. As a result, the 
value of this term is greatly increased compared to the stationary case.  Significant
increase (compared to the non-rotating case) also occurs for $C_{3}$-related term, suggesting
that modifications to the associated model coefficient might be required so that the model
attains the proper anisotropy levels. 
In accordance with the previous case study, the proposed model predicts smaller values for 
the normal component than Younis' formulation, achieving better agreement with the DNS in the near wall region 
as well as close to channel's centerline.   
\FloatBarrier

\subsection{Heated channel flow subjected to wall-normal rotation}

\begin{figure}[h!]
\flushleft
\subfloat[]
{\label{fig:RE_160_Ro004_WN_a}
\includegraphics[width=0.49\textwidth]{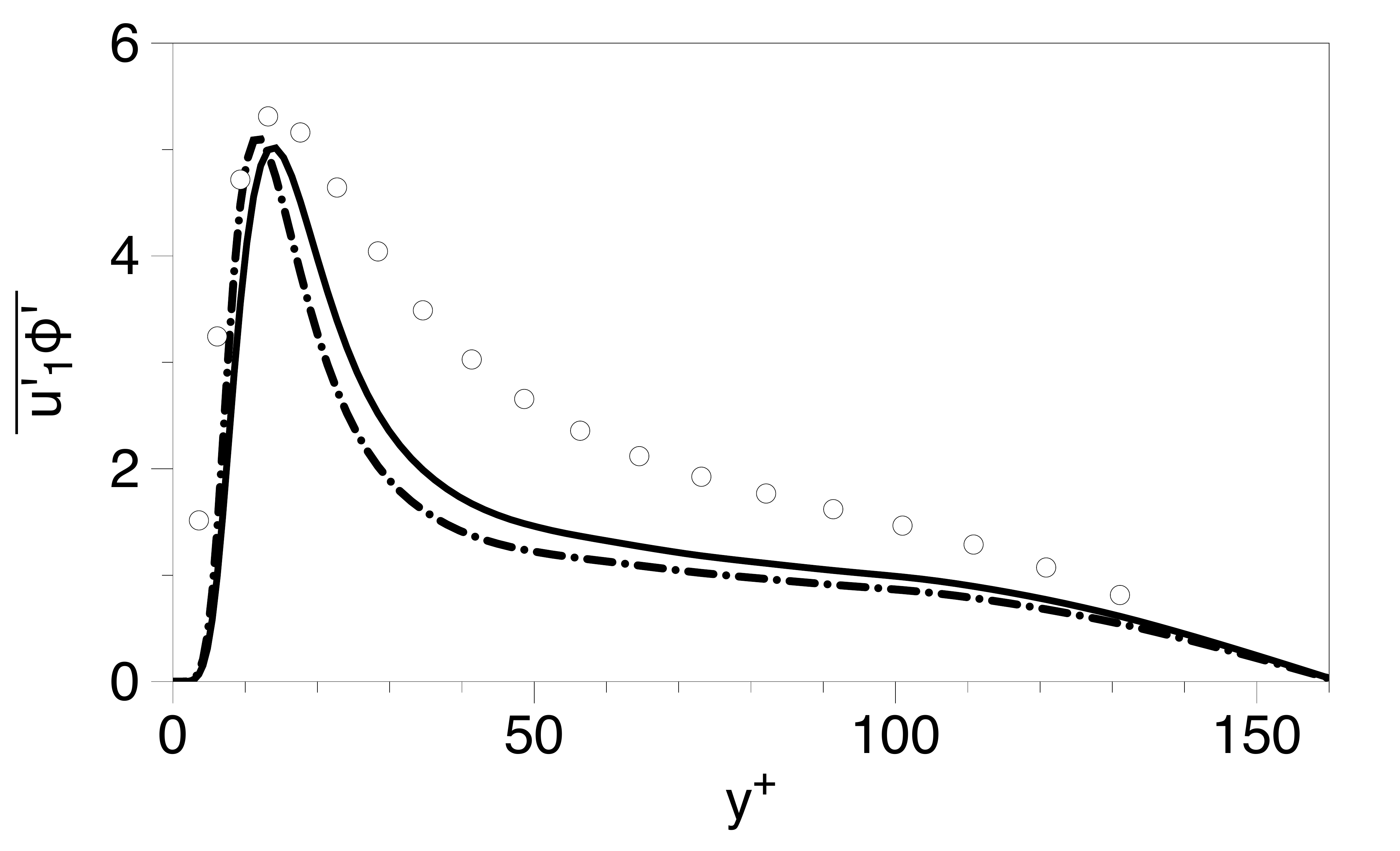}} 
\subfloat[]
{\label{fig:RE_160_Ro004_WN_b}
\includegraphics[width=0.49\textwidth]{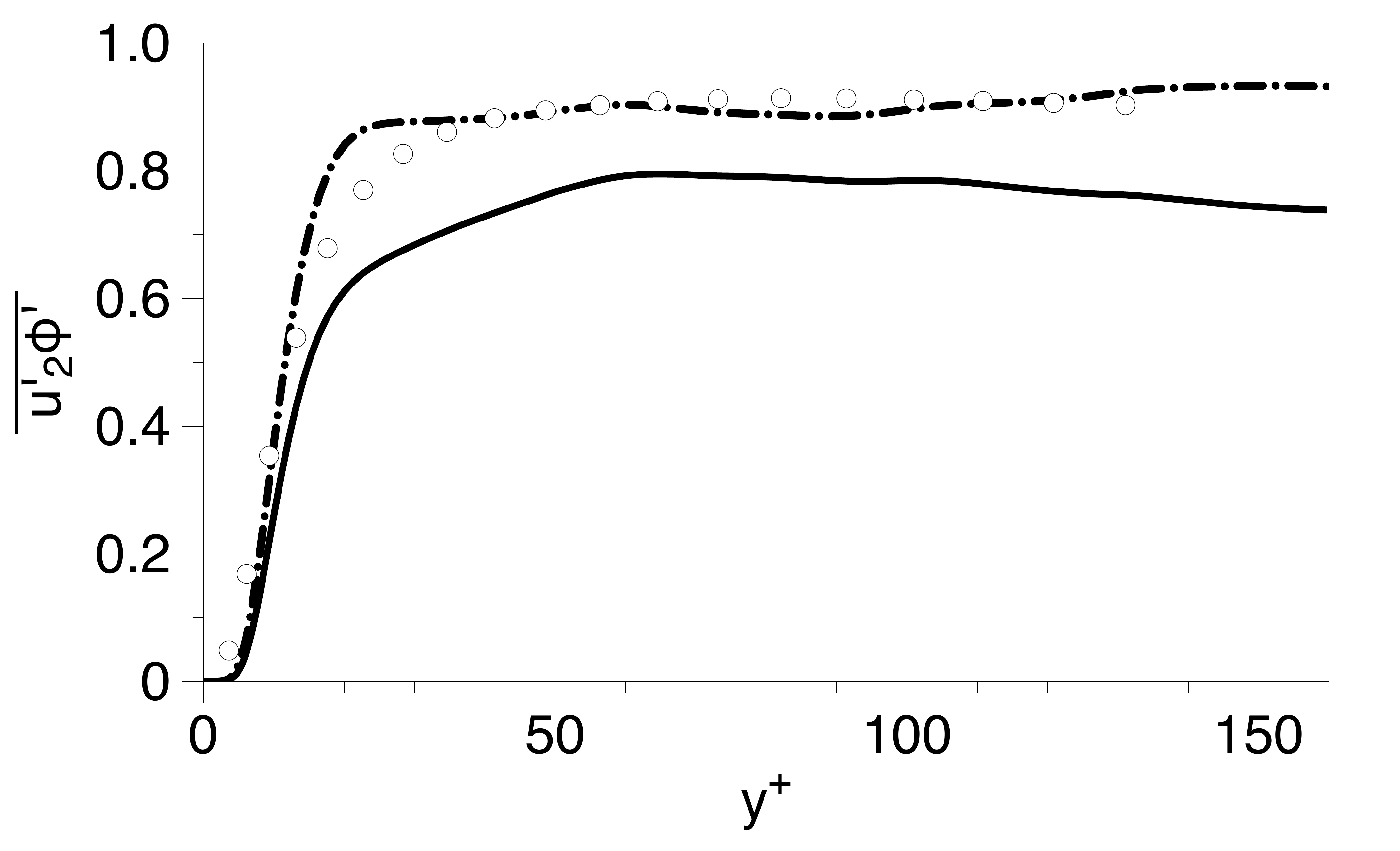}}\\
\caption{ Model predictions for the (a) streamwise and (b) wall-normal scalar flux components
for a heated channel flow at $Re_{\tau^{*}}=160$ that rotates around the wall-normal axis at
$Ro_{\tau}=0.04$. Solid line ($\solid$) denotes the proposed model and dash-dotted line 
($\chndot$) denotes Younis' model. Symbols denote the DNS results of  
El-Samni \& Kasagi \cite{ElSamni2000}. 
\label{fig:RE_160_Ro004_WN}
} 
\end{figure}

Here, the simulated flow field is rotated at a specified angular velocity around
the wall-normal axis, which induces a strong spanwise mean velocity that makes the absolute 
mean flow tilt towards to the spanwise direction.
The DNS results used for this case are those of El-Samni \& Kasagi \cite{ElSamni2000} at 
$Re_{\tau^{*}}= 160.1$ and $Ro_{\tau} = 0.04$.  
The profiles of the streamwise scalar-flux component are shown in figure~\ref{fig:RE_160_Ro004_WN_a}. 
The models provide  similar predictions, being able to capture accurately the near-wall peak 
while underestimating the value of this quantity in the remaining region.
As shown in figure~\ref{fig:RE_160_Ro004_WN_b}, Younis' model achieves very accurate predictions
for the normal component, while the present model underestimates the level of this quantity
outside the proximity of the wall boundary ($y^{+}>20$).
\FloatBarrier
\subsection{Heated channel flow subjected to spanwise rotation}

\begin{figure}[h!]
\flushleft
\subfloat[]
{\label{fig:RE_295_Ro25_SP_a}
\includegraphics[width=0.49\textwidth]{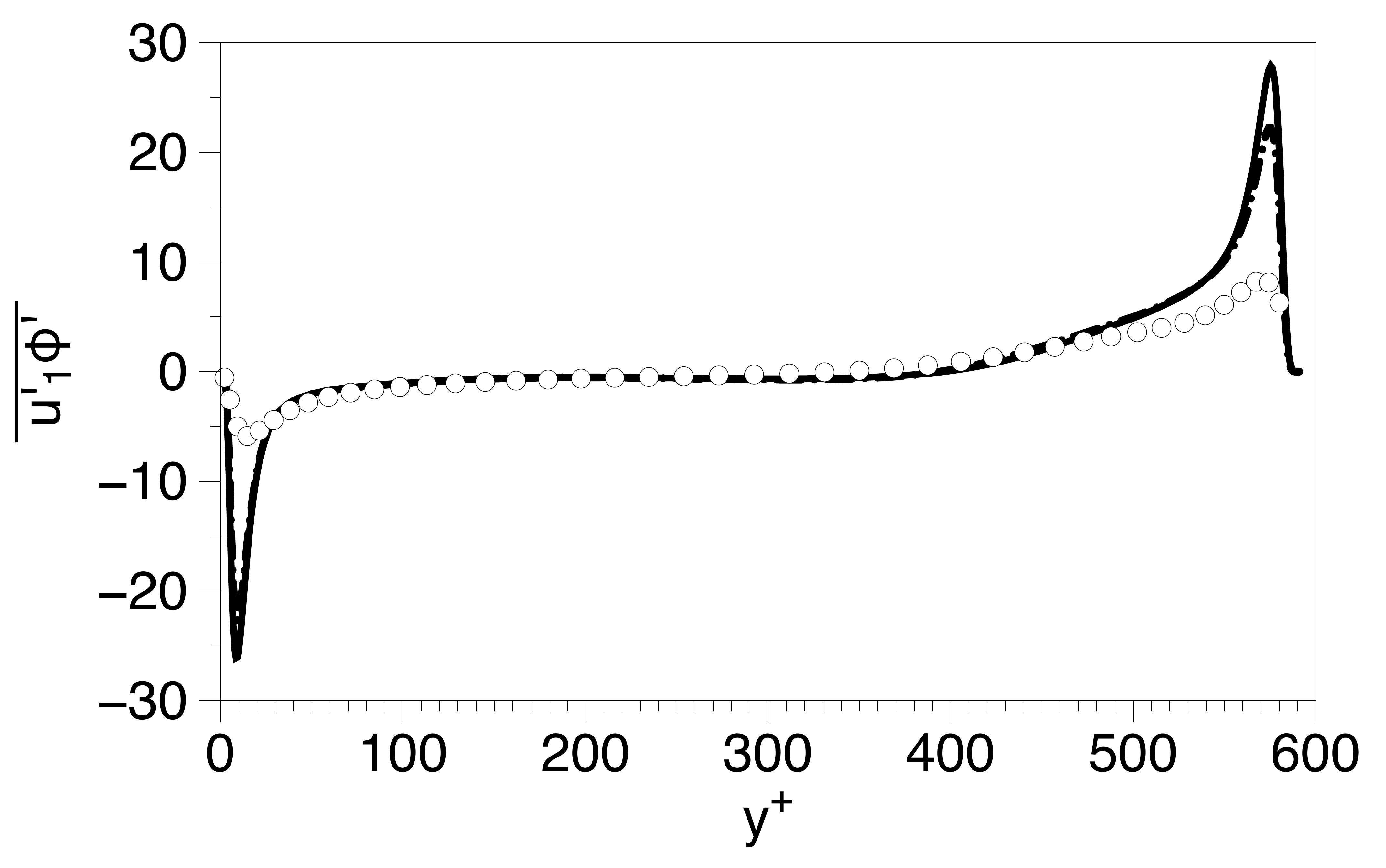}} 
\subfloat[]
{\label{fig:RE_295_Ro25_SP_b}
\includegraphics[width=0.49\textwidth]{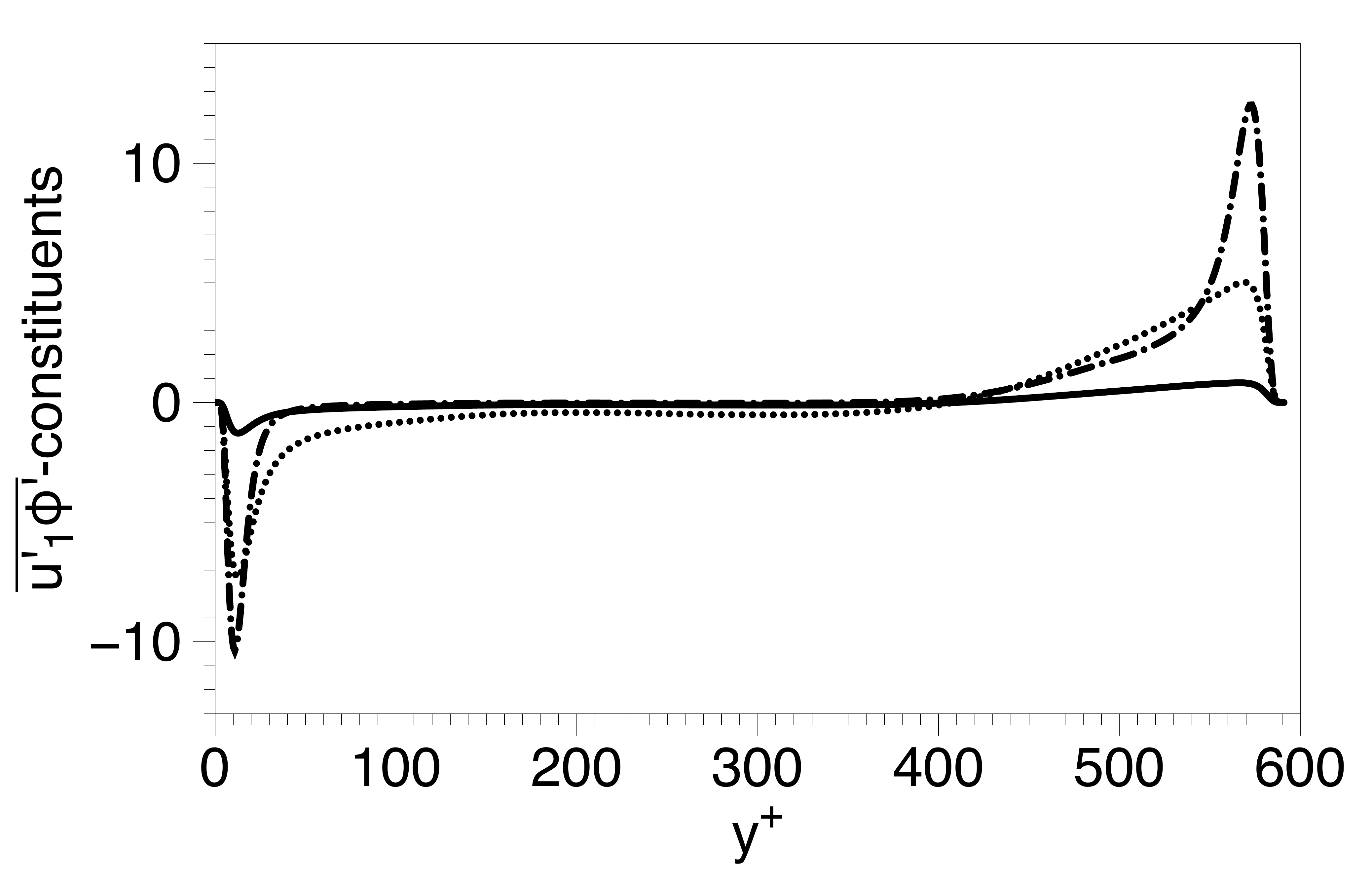}}\\
\caption{ (a) Model predictions for the streamwise scalar-flux component for a heated channel 
flow at $Re_{\tau^{*}}=295$ that rotates around the spanwise axis at $Ro_{\tau}=2.5$.
Solid line ($\solid$) denotes the proposed model and dash-dotted line ($\chndot$) denotes 
Younis' model. Symbols denote the DNS results of Wu \& Kasagi \cite{Wu2004}. 
(b) Constituents of the streamwise scalar-flux component according to equation~\eqref{u1phi_eq}. 
Solid line ($\solid$),  dashed-dotted line $(\chndot)$ and dotted line ($\dotted$) refer
to the terms associated with the $C_{2}$, $C_{3}$ and $C_{5}$ coefficients respectively.
\label{fig:RE_295_Ro25_SP}
} 
\end{figure}

\begin{figure}[h!]
\centering
\includegraphics[width=0.80\textwidth]{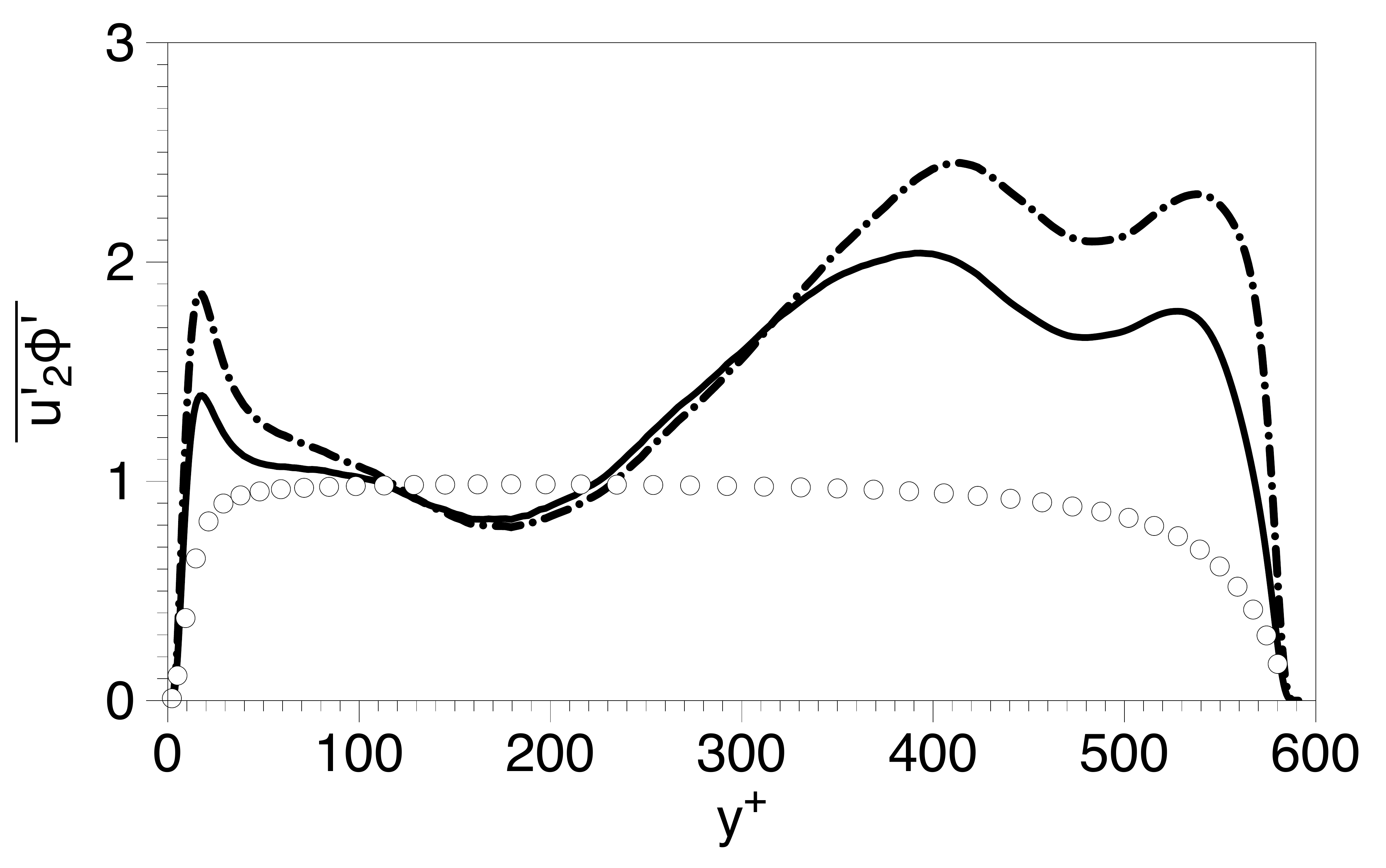}
\caption{ As in Figure~\ref{fig:RE_295_Ro25_SP_a}, but for the wall-normal component.
\label{fig:Q2_RE_295_Ro25_SP}
} 
\end{figure}

Next, we consider the presence of  Coriolis forces emerging from the spanwise rotation.
For this configuration, existing studies \cite{Grundestam2008, Xia2016} have shown that at 
moderate $Ro$ numbers, turbulence and especially the wall-normal fluctuations are augmented on the channel 
side where the system rotation is anti-cyclonic (called the unstable side), while they are 
damped on the channel side where the rotation is cyclonic (called the stable side). 
The scalar is kept at constant but different values at each wall, in accordance with the 
previous rotating cases.
Model predictions are compared with the DNS results of Wu \& Kasagi \cite{Wu2004} for
$Re_{\tau^{*}}=295.5$ and $Ro_{\tau}=2.5$. 
Figure~\ref{fig:RE_295_Ro25_SP_a} shows that both models severely overestimate the level 
of the peak magnitude in the near-wall region. Focusing on the present model, the strength 
of the individual terms that contribute to this quantity \eqref{u1phi_eq} is shown in 
figure~\ref{fig:RE_295_Ro25_SP_b}.
We observe the failure of the model in the near-wall region mainly because the $C_{3}$-related 
term becomes extremely high in the high-strain rate region, while it becomes trivial for 
$y^{+}>50$, yielding reasonable predictions of the model outside the buffer layer ($y^{+}>30$).
Consequently, modifications on the damping function $f_{C_{3}}$ should be applied to improve
model performance in this region.
Figure~\ref{fig:Q2_RE_295_Ro25_SP} reveals that both models are sensitized to the rotation-induced
asymmetry, with the proposed model being in closer agreement with the DNS results than Younis'
model. Note that equation~\eqref{u2phi_explicit_eq} indicates that the  $C_{3}$-related term does 
not contribute to the estimation, thus supporting the notion that modifications on the $f_{C_{3}}$ 
can improve model performance for this case. 
\FloatBarrier

\FloatBarrier

\section{ SUMMMARY AND CONCLUSIONS }\label{SUMMARY_AND_CONCLUSIONS}

In this study we have proposed an explicit algebraic closure for the turbulent scalar-flux
vector based on the general formulation of Younis, which is essentially an extended version of 
the model proposed by Abe \& Suga \eqref{abes_proposal} through the inclusion of an extra 
term involving products between the gradients of the mean velocity and scalar field. 
This closure consists of three terms, making it simpler than Younis' model and an elegant choice 
for use in general-purpose computational codes. It is also motivated by the ``effective-gradients"
hypothesis, which postulates that turbulence-turbulence interactions provide a gradient that 
acts supplementary to the mean shear, thus giving an alternative physical interpretation of
the proposal compared to other closures.
The resulting model explicitly depends on Coriolis effects in a similar manner to other models, 
which ensures its consistency with the principle of coordinate invariance.
To minimize model bias to inhomogeneous applications, the values of the model 
coefficients are determined based on existing LES predictions of homogeneous shear flows in 
the presence of arbitrary mean scalar gradients, while a simple damping function was applied
to account for the near-wall effects.
In order to test the quality of the proposed model, we have considered different types of heated 
channel flows, particularly Poiseuille and Couette flows, in both stationary and rotating 
frames. 
The proposed model performs well in Couette flows at different Reynolds numbers, showing
its sensitivity to the near-wall turbulent anisotropies. Regarding Poiseuille flows, 
good predictions are obtained for both flux components, with the model showing a mild tendency 
to overestimate the near-wall peak magnitude as the Reynolds number increases. Furthermore,
the proposed model captured the proper near-wall limit for the scalar ratio $R_{u'\phi'}$ for 
wall distances up to $y^{+} \approx 20$, significantly further than Younis' model ($y^{+} \approx 10$).   
In all cases, the proposed model achieved substantially better agreement with the DNS results than Younis' 
model for the normal component, while similar predictions between the models have been obtained 
for the streamwise component. 
The above non-rotating cases served as a guidance regarding the role of the different terms 
appearing in  the algebraic expressions, showing the ability of the proposal to capture the 
anisotropy between the different flux components.
The performance of the proposed model was further challenged in heated channel flows subjected to 
different modes of rotation, resulting in more complex flow configurations due to the 
emergence of secondary flows. 
Generally, the presented algebraic expression provides reasonable predictions for the normal
component. 
Under spanwise rotation though, the model fails in capturing the proper near-wall behavior 
(the same is also evident with Younis model) of the streamwise component, 
suggesting that further modifications on the damping functions should be investigated.   
Future work will focus on testing the proposed closure in additional cases involving frame-rotation,
in an attempt to further improve the estimation performance in the presence of Coriolis effects.
For that purpose, we intend to use DNS data in heated channel flows to quantify the response 
of passive scalar transport for a wider range of Rossby numbers.

\newpage 
\appendix
\section{Details regarding the calibration process.}\label{appendix} 

The model coefficients of the proposed algebraic closures are determined based on the LES data
of Kaltenbach et al. \cite{kaltenbach1994}, who investigated the turbulent transport of three 
passive species which have uniform gradients in either the normal, streamwise or spanwise direction.
All cases under consideration start from the same  shear parameter ratio 
$S^{*}_{o} = \bigg( \frac{ 2 S \kappa}{\epsilon} \bigg)_{o} = 5.04$ and the same Prandtl number $Pr=1.0$, in the initial absent of scalar
fluctuations. The mean flow configuration is expressed by
\begin{equation}
G_{ij} = S\delta_{i1}\delta_{j3}\,,\qquad \Lambda_{i} = \Lambda \delta_{i\alpha}\,,
\end{equation}
where greek index $\alpha$ denotes the direction of the mean scalar gradient. 
\begin{table}[h!]
\centering
\begin{tabular}{c c c c c c c c c c}
\hline
$St$     &  $R_{11}$   &  $R_{13}$  &   $R_{22}$   &   $R_{33}$  & $\epsilon$  &    $\Lambda_{i}$       &  $\overline{ u'_{1}\phi' } $  &  $ \overline{ u'_{2}\phi' } $   &   $ \overline{ u'_{3}\phi' } $      \\ \hline
8 		 &  4.98       &   -1.42    &    2.82      &    1.60     &  0.911      &  $( 1,0,0 )$		&  -10.9          			    &                  		          &       2.22       				    \\
10 		 &  6.32       &   -1.84    &    3.63      &    2.20     &  1.190      &  $( 1,0,0 )$		&  -13.9          			    &                  				  &       3.01           				\\
12 		 &  7.67       &   -2.27    &    4.71      &    3.12     &  1.600      &  $( 1,0,0 )$		&  -16.3          			    &                  			      &       4.08          				 \\
8 		 &  4.98       &   -1.42    &    2.82      &    1.60     &  0.911      &  $( 0,1,0 )$		&                 			    &     -3.84        				  &                      				\\
10 		 &  6.32       &   -1.84    &    3.63      &    2.20     &  1.190      &  $( 0,1,0 )$		&                 			    &     -4.62        				  &                     				 \\
12 		 &  7.67       &   -2.27    &    4.71      &    3.12     &  1.600      &  $( 0,1,0 )$		&                 			    &     -5.78        				  &                     				 \\         
8 		 &  4.98       &   -1.42    &    2.82      &    1.60     &  0.911      &  $( 0,0,1 )$		&    4.32         			    &                  				  &       -1.83          				\\
10 		 &  6.32       &   -1.84    &    3.63      &    2.20     &  1.190      &  $( 0,0,1 )$		&    5.42         			    &                 				  &       -2.47         				\\
12 		 &  7.67       &   -2.27    &    4.71      &    3.12     &  1.600      &  $( 0,0,1 )$		&    6.33         			    &                  				  &       -3.42         				 \\  \hline     
\end{tabular}
\caption{ Summary of the LES data used to determine model coefficients. Data are extracted 
at different total shear instants $St$ (where $t$ is time). }
\label{table:calibration_data}
\end{table}
\FloatBarrier

\newpage
\bibliography{references.bib}
\bibliographystyle{unsrt}
\end{document}